\documentclass[pra,aps,twocolumn,floatfix]{revtex4-1}
\usepackage[latin9]{inputenc}
\usepackage{graphicx}
\usepackage{amssymb}
\usepackage{amsmath}
\usepackage{amsfonts}
\usepackage{epsfig}
\usepackage{wrapfig}
\usepackage{refcount}
\usepackage{bm}
\usepackage{hyperref}
\usepackage{verbatim}
\usepackage[margin=1in,nofoot]{geometry}

\hypersetup{
    colorlinks=true,       
    linkcolor=cyan,          
    citecolor=magenta,        
    filecolor=magenta,      
    urlcolor=cyan,           
    runcolor=cyan
}

\newcommand {\be}{\begin{equation}}
\newcommand {\ee}{\end{equation}}
\newcommand {\bea}{\begin{eqnarray}}
\newcommand {\eea}{\end{eqnarray}}

\newcommand{\cl}{\mathcal}
\newcommand{\tr}{\mathrm{Tr}}

\usepackage[margin=1in,nofoot]{geometry}

\begin{document}

\title{Noise suppression via generalized-Markovian processes}

\author{Jeffrey Marshall}
\affiliation{Department of Physics and Astronomy, and Center for Quantum Information
Science \& Technology, University of Southern California, Los Angeles,
CA 90089-0484}

\author{Lorenzo Campos Venuti}

\affiliation{Department of Physics and Astronomy, and Center for Quantum Information
Science \& Technology, University of Southern California, Los Angeles,
CA 90089-0484}

\author{Paolo Zanardi}

\affiliation{Department of Physics and Astronomy, and Center for Quantum Information
Science \& Technology, University of Southern California, Los Angeles,
CA 90089-0484}

\begin{abstract}
It is by now well established that noise itself can be useful for performing quantum information processing tasks.
We present results which show how one can effectively reduce the error rate associated with a noisy quantum channel, by counteracting its detrimental effects with another form of noise.
In particular, we  consider the effect of adding on top of a purely Markovian (Lindblad) dynamics, a more general form of dissipation, which we refer to as generalized-Markovian noise. This noise has an associated memory kernel and the resulting dynamics  is described by an integro-differential equation.
The overall dynamics are characterized by decay rates which depend not only on the original dissipative time-scales, but also on the new integral kernel. We find that one can engineer this kernel such that the overall rate of decay is lowered by the addition of this noise term.
We illustrate this technique for the case where the bare noise is described by a dephasing Pauli channel. We analytically solve this model, and show that one can effectively double (or even triple) the length of the channel, whilst achieving the same fidelity, entanglement, and error threshold.
We numerically verify this scheme can also be used to protect against thermal Markovian noise (at non-zero temperature), which models spontaneous emission and excitation processes.
A physical interpretation of this scheme is discussed, whereby the added generalized-Markovian noise causes the system to become periodically decoupled from the background Markovian noise.

\end{abstract}
\maketitle
\section{Introduction}
Quantum systems interacting with an environment (open systems) are of increasing relevance for the understanding and practical application of quantum physics, in general. In particular, one of the biggest challenges in the experimental quantum computing community is designing devices which are robust against environmental noise \cite{Landauer:95,Unruh:1995fk}. Combating such noise has become itself a field of research, and has led to the development of pioneering techniques, broadly referred to as quantum error correction or error suppression \cite{Lidar-Brun:book}.

Recently however, it has become clear that noise itself can in fact be exploited to the end of performing quantum information processing (QIP) tasks. The early work in this area focused on encoding entangled states \cite{Kraus-prep} and even the output of a computation \cite{verstraete2009quantum} in the steady state of a dissipative dynamics. Since then other results have appeared which show how one can enact simulations of quantum systems, both open and closed \cite{barreiro2011open,PhysRevA.83.062317,markov_simulation_barthel,zanardi-dissipation-2014,dissi_2nd_order:Zanardi2016}, and even perform general computations (robust to certain types of error) in the presence of strong dissipation \cite{DGM}.

Motivated by the recent progress in simulating non-Markov systems \cite{NM_simulation_12,NM_simulation_15,markovianity_tunable,NM_sweke,simulation_disi15}, we introduce a reservoir engineering
technique \cite{res_eng_poyatos,PhysRevLett.86.4988,Bellomo_structured_env,
PhysRevA.81.062119,PhysRevA.81.062306,PhysRevA.81.052330,man:2015_cavity} whereby so-called generalized-Markovian dissipative processes (studied variously by e.g. \cite{ShabaniLidar:05,vacchiniME,semi_markov_pra})  can be exploited to the end of reducing the rate at which errors accumulate over a dissipative Markovian evolution.
We will show that upon adding generalized-Markovian noise on top of an assumed background Markovian channel, the rate at which the system approaches the steady state can be reduced; that is, it will take longer for the system to relax to the steady state, and one can for example, preserve quantum information encoded in arbitrary states for longer times. 

The paper is organized as follows: We will first set-up and define the general class of noisy systems we will be considering, before outlining our error suppression technique itself. Following this we provide physically motivated examples which illustrates this method for evolution over a noisy Pauli channel, and also the case of thermal noise.
We provide some analytic solutions to these models, and also numerically quantify the success of the scheme in these situations. We finish with a general discussion setting our results in a broader picture.

\section{Set-up}
We assume we have some noisy `background' quantum channel which is to a good approximation
described by a time-independent master equation of the Lindblad type (i.e. the channel is Markovian). We write 
\be
\dot \rho(t) = \cl{L}_0 [\rho(t)],
\label{mainEq}
\ee 
where $\cl{L}_0$ is a generator of Markovian dynamics \cite{Lindblad:76} \footnote{$\dot X := dX/dt$}. We will assume throughout the dimension of the Hilbert space of the system is finite.

It is convenient to introduce the spectral (Jordan) decomposition of $ \cl{L}_0 $ \cite{Kato:book}:
\begin{equation}
\cl{L}_0 = \sum_i \lambda_i\cl{P}_i + \cl{D}_i. \label{eq:L0_spectral}
\end{equation}
The eigen-projectors $\cl{P}_i$ ($\sum \cl{P}_i=1$) and eigen-nilpotents $\cl{D}_i$ satisfy: $\cl{P}_i\cl{P}_j = \delta_{i,j}\cl{P}_i,\, \cl{D}_i\cl{P}_j = \cl{P}_j\cl{D}_i = \delta_{i,j}\cl{D}_i$ [with $\cl{D}_i\cl{D}_j = \delta_{i,j}\cl{D}_i^2$]. Also, there is an integer $m_i \ge 0$ such that $\cl{D}^{m_i}=0$ [and $\cl{D}_i^{m_i - 1}\neq 0$, when $m_i>0$].

If $\cl{L}_0$ is of the Lindblad form, we also have that $\text{Re}\,\lambda_i \le 0$, and there is guaranteed to be at least one zero eigenvalue (with no eigen-nilpotent part) (see e.g., \cite{wolf_quantum_2012,venuti_adiabaticity_2016}). 
The zero eigenvalue states span the so-called steady state space. If the non-zero eigenvalues have negative real parts (i.e., not purely imaginary), the steady state manifold is attractive and the evolution over infinite time brings any initial state to the steady state space.

Using Eq.~(\ref{eq:L0_spectral}) we can write the evolution (super) operator, $\Phi_0(t):=e^{t\cl{L}_0}$, and the resolvent, $R(z):=(z-\cl{L}_0)^{-1}$ as

\begin{equation}
\label{spectral-evolution}
\Phi_0(t) = \sum_i \left( \cl{P}_i + \sum_{k=1}^{m_i-1}\frac{t^k \cl{D}_i^k}{k!}\right)e^{\lambda_i t} 
\end{equation}
and
\begin{equation}
\label{resolvent}
R(z) = \sum_i \left(\frac{\cl{P}_i}{z - \lambda_i} + \sum_{k=1}^{m_i-1}\frac{D_i^k}{(z - \lambda_i)^{k+1}}\right).
\end{equation}

From Eq.~(\ref{spectral-evolution}) the decay rates of the channel $\Phi_0(t)$ are determined by the real part of the eigenvalues, in particular,  $\tau_{0,i}^{-1}:= |\text{Re}\lambda_i|$ defines the decay time in the  $i$-th block. Our goal is to engineer a channel as close as possible to the identity channel (given the above fixed background). As time increases, a channel of the form of $\Phi_0(t)$ departs (in a possibly non-monotonic way) from the ideal channel at $t=0$. In this sense we see that it is the short time-dynamics that are important for our purposes. In other words the behavior we are interested in is characterized by the \emph{shortest} time scale $\tau_0 = \mathrm{min}_i \, \tau_{0,i}$. This is to be contrasted with another typical situation, where one is interested in the approach to the steady state which is instead dictated  by the longest time scale.

To quantify how much a channel $\Phi$ departs from the ideal one, we use a fidelity based measure: given a quantum channel $\Phi$, we define the \textit{minimum channel fidelity} $f$ as
\be
f(\Phi):=\min_\rho F(\rho, \Phi(\rho)),
\label{channel-fidelity}
\ee
where $F(\rho,\sigma) :=(\tr \left (\sqrt{ \sqrt{\rho} \sigma \sqrt{\rho}} \right ))^2$  is the fidelity between the states   $\rho,\, \sigma$. This essentially tells us the worst case performance of this channel over all states. 
Note, by convexity this minimization can be carried out over pure states.

In general, one would like to set some minimum error threshold $\epsilon$ such that only channels satisfying, $f(\Phi) \ge 1- \epsilon$, for some $\epsilon > 0$ are tolerated.
However, given some fixed background channel (e.g., as above), finding necessary and sufficient conditions to increase $f$ is a very complicated task. In the next section, we will show how one can decrease the decay rates $\tau_{0,i}^{-1}$, thus improving the quality of the channel as a whole. This, in particular, will increase  the minimum channel fidelity.

\section{Methods}
On top of a Markovian,  dissipative background we now add, at the master equation level, a secondary form of noise, which we refer to as generalized-Markovian noise. The dynamics are now given by the following master equation
\be
\label{masterEq}
\dot \rho(t) = \cl{L}_0\rho(t) + \cl{L}_1 \int_0^t k(t-t')\rho(t')dt',
\ee
where $\cl{L}_1$ is time-independent, and also of the Lindblad form. We refer to $k(t)$ as the memory kernel. For convenience we also define $\cl{K}$ such that $\cl{K}X = \int_0^t k(t-t')X(t')dt'$.

Purely Markovian (Lindbladian) dynamics are recovered if the kernel is of the form $k(t-t') = \delta(t-t')$. 
It is known (see e.g., Ref.~\cite{hazards}) that it is possible to find kernels such that the resulting evolution operator is not completely positive (CP). Here we require that Eq.~(\ref{masterEq}) is such that the generated dynamics are CP for all $t\ge0$ . The examples we provide below all fulfill this important criterion. 
On physical grounds we also assume that $\cl{L}_0$ and  $\cl{L}_1\cl{K}$ originate from separate processes, and therefore we require that $\cl{L}_1\cl{K}$ must also generate a genuine quantum (CP) map alone. With this constraint  we are not allowed to fulfill our goal by simply taking, e.g., $\cl{L}_1 = -\cl{L}_0$, with $k(t)=\delta(t)$.

As is well known, if the Lindblad operators for $\cl{L}_1$ are self-adjoint,  a master equation of the form of Eq.~(\ref{masterEq}) can be obtained by coupling a suitable Hamiltonian to a (classical) stochastic noise term (see for example Ref.~\cite{PhysRevA.70.010304}). In this approach the kernel $k(t)$ originates as the autocorrelation function of the classical, stochastic field. We provide a brief reminder of this approach in  Appendix \ref{sec:appendix_ME}.

We solve Eq.~(\ref{masterEq}) by taking the Laplace transform:
\begin{equation}
\label{general-laplace}
\tilde{\rho}(s) = \frac{1}{s - \cl{L}_0 - \tilde{k}(s)\cl{L}_1 }\rho(0)=:\tilde{\Phi}(s)\rho(0)
\end{equation}
where 
\begin{equation}
\tilde{f}(s) = L[f(t)](s) = \int_0^\infty e^{-s t } f(t) dt
\end{equation}

is the notation for the Laplace transformation of $f$.

At this point we make the important assumption that  $\cl{L}_0$ and $\cl{L}_1$ have the same spectral decomposition. Note that this is in principle not a necessary requirement for the success of our scheme (e.g. as will be shown in Sect.~\ref{sec:thermal}), however it provides a useful insight into its mechanism.
 In this case using Eq.~(\ref{resolvent}) we can write
\begin{equation}
\label{laplace-solution}
\begin{split}
\tilde{\Phi}(s) = \sum_i \left[\tilde{\Lambda}_i(s)\cl{P}_i\ + \sum_{k=1}^{m_i-1} \left[ \tilde{\Lambda}_i(s) \right]^{k+1}D_i^k \right],
\end{split}
\end{equation}
with 
\begin{equation}
\tilde{\Lambda}_i(s) = \frac{1}{s - \lambda_i - \tilde{k}(s)\mu_i}, \label{eq:lambda_tilde}
\end{equation}
 where $\lambda_i,\,\mu_i$ are the eigenvalues of $\cl{L}_0,\,\cl{L}_1$ respectively associated with the $i$-th eigenspace.
The evolution operator is then given by $\Phi(t) = L^{-1}[\tilde{\Phi}(s)](t)$.

Consider for example the case where $\tilde k(s)=p(s)/q(s)$ is a rational function with  polynomials $p,q$. This corresponds to a large class of kernels which are (finite) linear combinations of functions of the form $t^n e^{a t}$ for complex $a$ and integer $n$.
In this case  one can write $\tilde \Lambda_i(s) = P_i(s)/Q_i(s)$ (with no common roots between $P_i$ and $Q_i$). Note by construction we have  $\text{deg}(P_i) < \text{deg}(Q_i)$, so one can always write $\tilde{\Lambda}_i(s)$ as a partial fraction decomposition
\begin{equation}
\tilde{\Lambda}_i(s) = \sum_j \sum_{n_j=1}^{N_j^{(i)}} \frac{c_{n_j}^{(i)}}{\left(s - s^{(i)}_j\right)^{n_j}}, \label{eq:PFD}
\end{equation}
where the roots $s_j^{(i)}$ of $Q_i(s)$ occur with multiplicity $N_j^{(i)}$, and the $c$ are constants.
Laplace transforming Eq.~(\ref{eq:PFD}) back we obtain:
\be
\Lambda_i(t) = \sum_j \left(\sum_{n_j=1}^{N_j^{(i)}} c^{(i)}_{n_j} \frac{t^{n_j -1 }}{(n_j-1) !}\right)e^{s^{(i)}_j t}.
\label{laplace-sol}
\ee
This function, in absence of the nilpotent terms, completely specifies the full map $\Phi(t)$. 
 
The real part of the roots $s^{(i)}_j$ therefore determine the rate of decay of the system.
These roots will depend not only on the eigenvalues $\lambda_i$, but also on the specific nature of the integral kernel $k$. The key observation we make is that for certain choices of $k$, the decay rate of the `combined' system can in fact be lower than that of the original `background' system.

In the $i$-th eigen-space, for $k(t)=0$, the decay is simply of the form $\Lambda_i(t) = e^{\lambda_i t}$. We see that if we can guarantee $|\text{Re}(s_j^{(i)})| < |\text{Re}(\lambda_i)|$, $\forall j$, then the rate of decay associated with this subspace will have effectively been reduced.
This is equivalent to $\tau_{i}^{-1} < \tau_{0,i}^{-1}$, where $\tau_{i}^{-1} := \text{max}_j |\text{Re}(s_j^{(i)})|$.
This can therefore result in an increase in the minimum channel fidelity over some fixed evolution time (as will be illustrated below).

We would like to remark here that if we set $k(t) = \delta(t)$, then under the same conditions as above it is not possible to reduce the decay rates $\tau_{0,i}^{-1}$. The non-trivial form of the memory kernel $k$ is completely central to this technique.

We now provide some examples, to illustrate our scheme.

\section{Example: Pauli Channel}

We consider the dynamics of an $N$ qubit  generalization of the standard single qubit Pauli channel \footnote{One can think of this as a model for a classically correlated noisy channel (if an error occurs, it occurs to all qubits simultaneously). See e.g.~Ref.~\cite{dynamicalMemory} for a two qubit version.}. 
Note, we take arbitrary $N$ only for generality, and that in practice, one is limited to $N=1,2$
(since otherwise more than 2-body couplings would be required, which is experimentally challenging).
We mention that this is an important class of noise since it is known that quantum error correction techniques can correct against arbitrary errors given the ability to correct against such dephasing errors \cite{Lidar-Brun:book}.

With this in mind, we take our background Markovian channel to be dephasing in the $k$-direction (where $k = 1,2,3$), via the Markovian generator
\begin{equation}
\mathcal{L}_0[X]  = \gamma (A_k X A_k - X),
\label{lindblad_L0}
\end{equation}
where $A_{k} =  \sigma_k^{\otimes N}$, $\sigma_k$ is the $k$-th Pauli matrix \footnote{We will occasionally make use of the spectral decomposition of \unexpanded{$\sigma_3 = |1\rangle \langle 1| - |0\rangle \langle 0|$}. Throughout, when we write \unexpanded{$|0\rangle,|1\rangle$}, it is in this $z$-eigenbasis.} and $\gamma>0$. The solution of this dynamics is given by the following quantum map
\begin{equation}
\Phi_0(t) [X] = (1-p_0(t))X + p_0(t)\, A_k X A_k,
\label{dephasing}
\end{equation}
where $p_0(t) = \frac{1}{2}(1 - e^{-2 \gamma t})$ is the probability of dephasing (i.e., with probability $p_0(t)$, the state $X$ will become $A_k X A_k$).
The minimum fidelity of this channel is $f_0(t):=f(\Phi_0(t)) = \frac{1}{2}(1+e^{-t/\tau_0})$, with the associated decay rate $\tau_0^{-1} = 2\gamma$.

In this case, there are no eigen-nilpotents, and one can write the spectral projection as
\begin{equation}
\label{spectral}
\cl{L}_0 = \sum_{\bar n} \lambda_{\bar n}\cl{P}_{\bar n} 
\end{equation}
where the sum is over all strings $\bar n = (n_1,\dots ,n_N)$, with $n_i \in \{0,1,2,3\}$.
The projectors are given by (see Appendix \ref{pauli-deriv})
 \be 
 \label{projectors}
 \cl{P}_{\bar n}(X) = \frac{1}{2^N}\text{Tr}(X \sigma_{\bar n})\sigma_{\bar n}, \ee 
 with $\sigma_{\bar n} := \otimes_{i=1}^N \sigma_{n_i}$ (and $\sigma_{0}=\mathbb{I}$, the 2$\times$2 identity matrix), while the eigenvalues are either 0, or $-2\gamma$.
The evolution operator can therefore  be written  as
$\Phi_0(t) = \sum_{\bar n} e^{\lambda_{\bar n} t}\cl{P}_{\bar n}$.

We define the projection to the steady state of the dynamics
(i.e. the infinite time limit of the evolution) as
\begin{equation}
\label{ss_proj}
P_0 := \lim_{t\rightarrow \infty} \Phi_0(t) = \sum_{\bar{n} : \lambda_{\bar n}=0}\cl{P}_{\bar n}.
\end{equation}
Note that for all quantum states $\rho$, the corresponding state $P_0\rho$
is steady in the sense that it does not evolve under $\cl{L}_0$; one can check (e.g.
using Eq.~(\ref{spectral})) that $\cl{L}_0 P_0 = 0$. We will exploit this in our scheme, as will be seen more explicitly below.

\subsection{Purely decaying noise}

To this background channel, we add generalized-Markovian noise as described above. We take $\cl{L}_1 \propto \cl{L}_0$ (i.e., equal up to a positive constant). We first take the memory kernel to be of the form $k(t-t') = B^2 e^{-|t-t'|/\tau_{k}}$ $\implies \tilde{k}(s) =B^2 \frac{1}{s+\tau_{k}^{-1}}$ (note, $\tau_{k} > 0$). One can think of $\tau_{k}$ as the characteristic time over which the memory associated with the added noise persists. We will absorb the coupling constant of $\cl{L}_1$ into the kernel strength $B$, to avoid introducing a redundant parameter.

For the eigenvalue $-2\gamma$, Eq.~(\ref{eq:lambda_tilde}) gives
\begin{equation}
\label{roots}
\tilde{\Lambda}(s) = \frac{1}{s + 2\gamma+ \frac{B^2}{s + \tau_{k}^{-1}}} = \frac{c_+}{s-s_+}+\frac{c_-}{s-s_-},
\end{equation}
for constants given by
\begin{eqnarray}
c_\pm & =& \frac{1}{2}\left(1 \pm i \frac{ \tau_{k}/\tau_{0} - 1}{2 \omega \tau_{k}}\right) \\
s_\pm &=&  -\tau^{-1} \pm i\, \omega \\
\omega &=& \sqrt{B^2 -\left(\frac{1}{2 \tau_{0}}- \frac{1}{2 \tau_{k}}\right)^2}.
\end{eqnarray}

Taking the inverse Laplace transform of Eq.~(\ref{roots}) we obtain
\be
\label{eq-lambda}
\Lambda(t) = e^{-t/\tau}\frac{\cos (\omega t + \phi)}{\cos \phi},
\ee
where the new decay rate is $\tau^{-1} = \frac{1}{2}(\tau_0^{-1} + \tau_{k}^{-1})$  and $\phi$ satisfies $\cos \phi =  2\omega \tau_{k}/ \sqrt{(2\omega \tau_{k} )^2+(\tau_{k}/\tau_0 - 1)^2 }$. Note that for $\omega =0$ the solution is slightly different (see Appendix \ref{sec:deriv} for more details).
Note that if $B=0$ (background channel alone), we recover $\Lambda(t) = e^{-t/\tau_0}$. 

The solution of this dynamics is therefore 
\begin{equation}
\label{newMap}
\Phi(t)[X] = (1-p(t)) X + p(t)\, A_k X A_k,
\end{equation}
where $p(t) = \frac{1}{2}(1-\Lambda(t))$. 


We show in the Appendix \ref{sec:cp} that $0 \le p(t) \le 1$  for all values of the parameters and $t\ge0$, i.e., this indeed generates a CP map. However we will focus on the case where $\omega \in \mathbb{R}_{+}$ corresponding to the condition  $2|B| > |1/\tau_0 - 1/\tau_k|$.

\begin{figure}[h]
\begin{centering}
\includegraphics[scale=0.4]{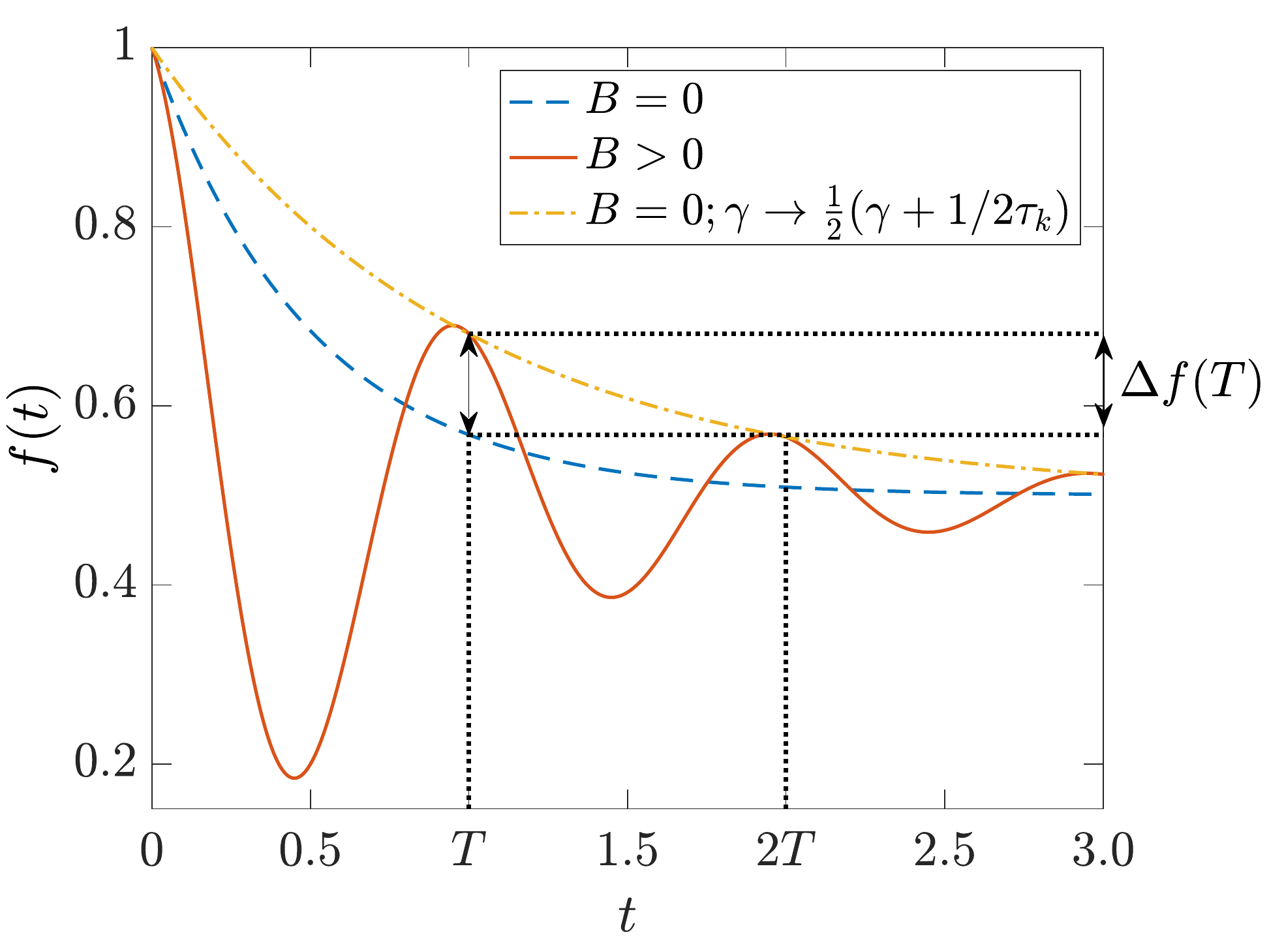}
\end{centering}
\vspace{-4mm}
\caption{Minimum channel fidelity $f(t)=\frac{1}{2}(1+\Lambda(t))$, as a function of time for the purely Markov (background channel) evolution (blue/dash, $B=0$), and for the combined system aided by the generalized-Markovian noise (red/solid, $B>0$). We also plot the fidelity of the background channel, with $\gamma$ rescaled to $\frac{1}{2}(\gamma + 1/2\tau_k)$ (yellow/dot-dash), which intersects the $B>0$ curve at times $t=\frac{2}{\omega}(\pi n - \phi)$and $2\pi n/ \omega$.
Here, $\gamma=1, \tau_{k} = 25$, and we set $B$ such that $\omega T = 2\pi$ (for $T=1$). Note, the axis of dephasing is not important here; $f$ is the same for each direction (for fixed parameters). Time is measured in units of $1/\gamma$.}
\label{fidelity-fig}
\end{figure}

We note, importantly, that the decay rate of the new system, $\tau^{-1}$, can in fact be less than the decay rate for the original (Markov) system alone, $\tau_0^{-1}$. 
This occurs when $\tau_{k} > \tau_0$, so that $\tau^{-1} < \tau_0^{-1} $. 
When this is the case, we find for certain times along the evolution that $p(t) < p_0(t)$, i.e. the probability of a dephasing error occurring is reduced. This is equivalent to an increase in the minimum channel fidelity $f$, see Fig.~\ref{fidelity-fig}.

From Eq.~(\ref{eq-lambda}), for times $t_n = 2 \pi n / \omega $ or $t_n = (2\pi n - 2\phi)/\omega$ ($n=1,2,\dots$), $\Lambda(t_n)= e^{-t_n/\tau}$ and the evolution is the same as that generated by $\cl{L}_0$, but with $\tau_0$ being replaced by $\tau$. For example, the limit $\tau_{k}\rightarrow \infty$ is equivalent to replacing $\gamma$ with $\gamma/2$ (and evolving for time $t_n$).  In other words we are able to change (e.g.~increase) the coherence time of the channel without changing any other of its properties.

As an example of a direct application of this, if one has some fixed length quantum channel of the form Eq.~(\ref{dephasing}), i.e. $t=T$ is fixed (equivalently, $p_0$ is fixed), we have shown that upon the addition of this generalized-Markovian noise to the system, the new channel will have error probability $p = \frac{1}{2}(1-\Lambda(T))$. If we pick $\tau_{k} > \tau_{0}$, i.e., the kernel decay time is longer than the bare channel decay time, upon chosing $B$ such that $\omega T=2\pi$, then we have
\be
p  = \frac{1}{2}(1-e^{-T/\tau}) < \frac{1}{2}(1-e^{-T/\tau_0}) = p_0
\ee
or equivalently, $f > f_0$. Note that since our map is CP for all parameter choices, we can always find such a choice for $B$.

In Fig.~\ref{fidelity-fig} we give an explicit example of this. We plot the fidelity over the dynamical evolution of the background channel, and also the combined channel. We see that along certain points of the evolution, the fidelity of the combined system surpasses that of the background [in the figure $\Delta f(T) \approx 0.1$].

Fig.~\ref{fidelity-fig} also shows that the minimum channel fidelity of the background channel (blue) at time $t=T$, is (approximately) equal to the fidelity of the combined channel (red) at time $t=2T$. In other words adding a noise term with a non-trivial kernel can be beneficial for performing quantum information processing tasks, for example, by allowing quantum states to be stored as a memory for a longer time (in this case, for nearly twice as long, whilst achieving the same minimum channel fidelity).

\begin{figure}[h]
\begin{center}
\includegraphics[width=0.45\columnwidth]{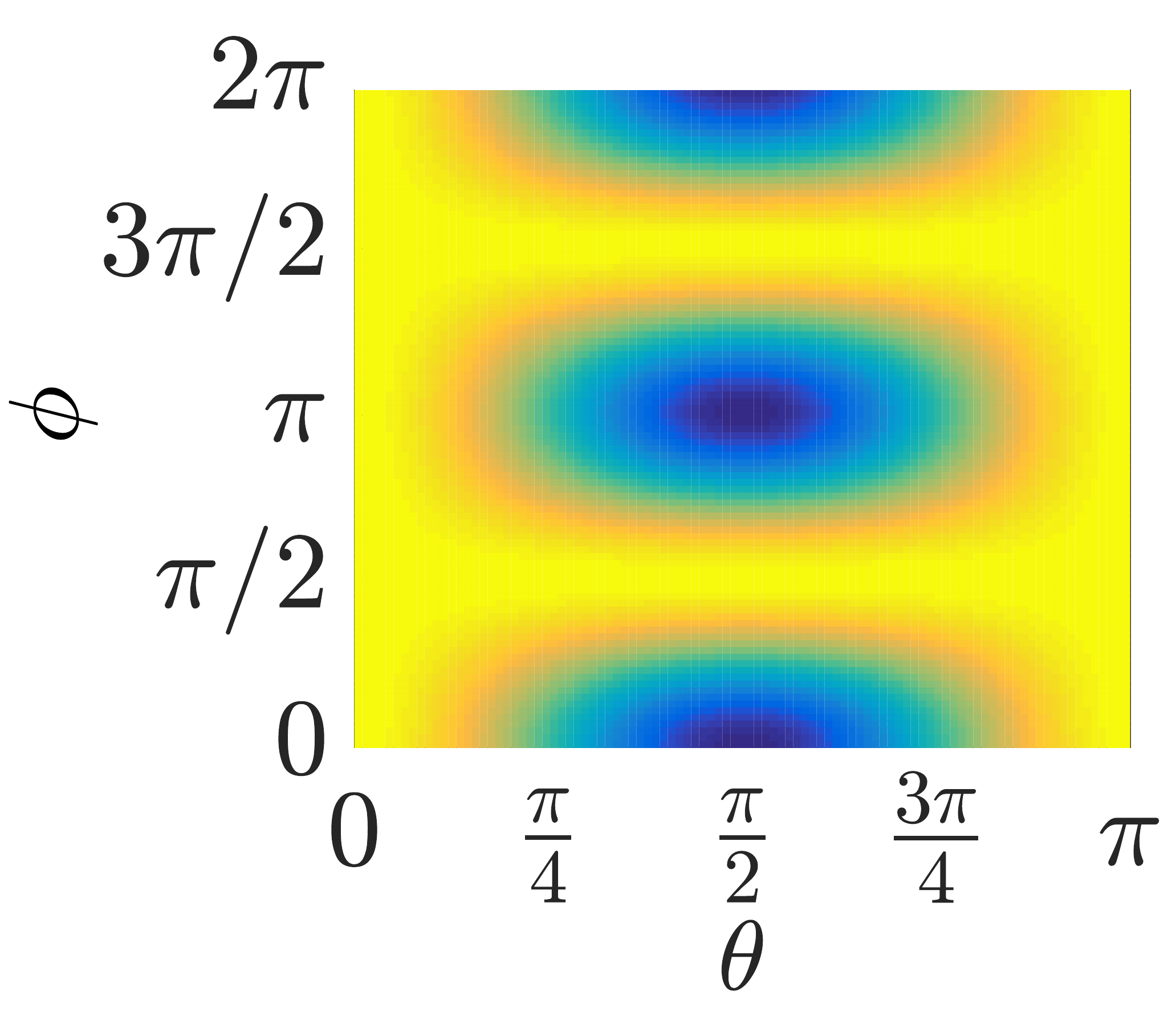}
\includegraphics[width=0.52\columnwidth]{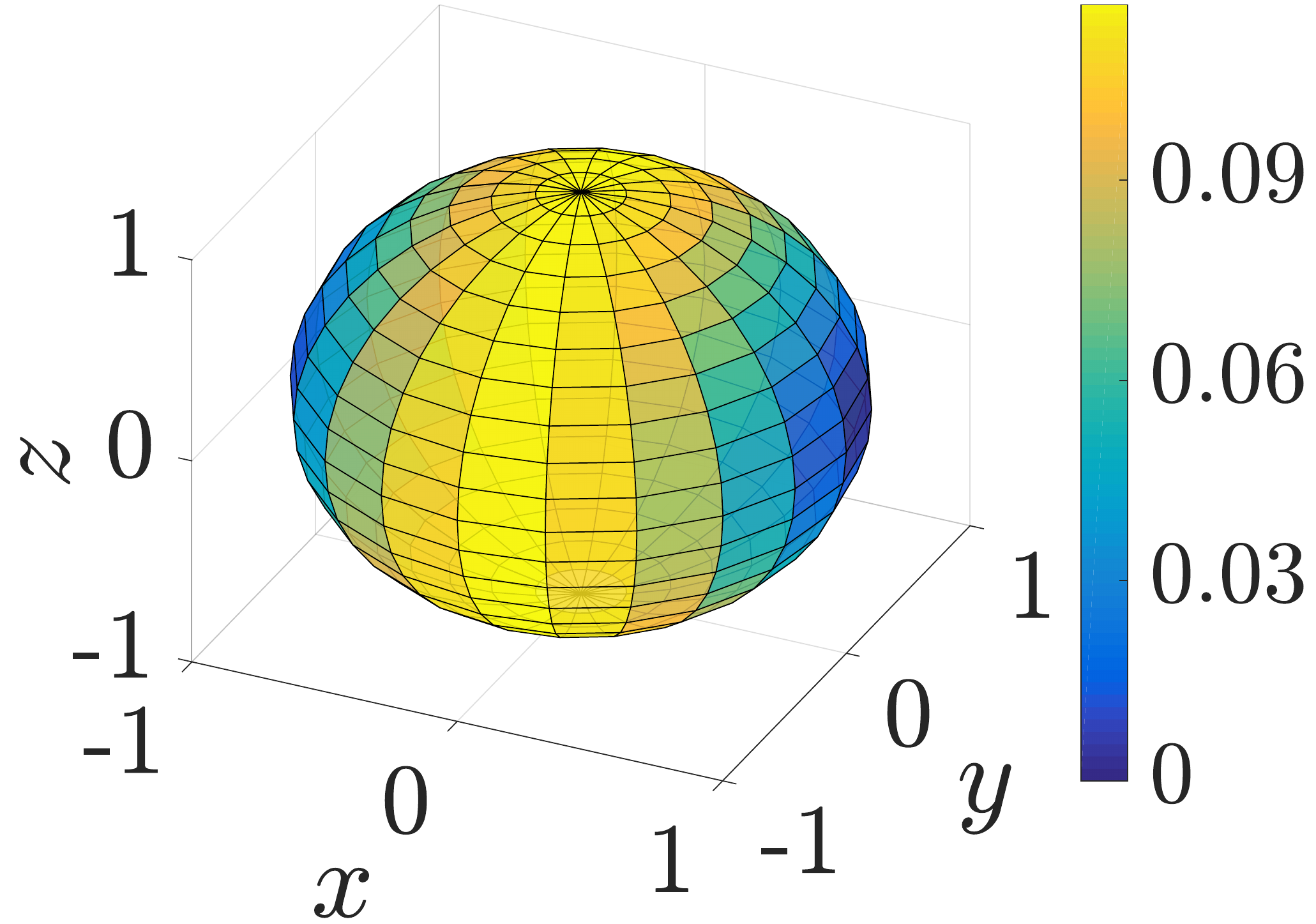}
\caption{Difference in the fidelity $\Delta F := F-F_0$ for all pure states (single qubit), where $F_0$ is the fidelity of the `background' channel, and $F$ the fidelity the `combined' channel, at time $t=T$ [cf. Fig.~\ref{fidelity-fig}]. The background channel here corresponds to dephasing in the $x$ direction.
(Left) $\theta,\phi$ axes correspond to positions on the Bloch sphere for a single qubit: $|\psi\rangle = \cos (\theta/2)|0\rangle + e^{i\phi}\sin (\theta/2)|1\rangle$. (Right) Spherical plot of $\Delta F$ (i.e. on the Bloch sphere).
We see for all non-stationary initial states, the fidelity increases. 
Parameters: $T=1,\gamma=1$ (hence $p_0 \approx 43\%$), and $\tau_{k} = 25$, with $B$ chosen so that $\omega T = 2\pi$ (hence $p \approx 32\%$).} 
\label{fig:contour}
\end{center}
\end{figure}

In Fig.~\ref{fig:contour}, we plot the difference in fidelity $\Delta F$ between the background channel ($x$-dephasing), and the channel assisted by the generalized-Markovian noise, for all single qubit pure states, at a fixed time $t=T$ (that is, we map $\Delta F$ at some fixed instance in time of the evolution, where each initial state is a single qubit pure state, as defined by the coordinates in the figure). 
This shows that the fidelity always increases apart from the steady states $|+\rangle,|-\rangle$ of the dynamics (which have maximal fidelity of 1 by definition, under both channels). For our choice of parameters, the error probability decreases from $p_0\approx 43\%$ to $p \approx 32\%$ (i.e. the probability of an error occurring over the channel is reduced by more than 10$\%$).

We also consider the effect of sending a single qubit of an entangled pair down the channel Eq.~(\ref{newMap}). 
 We quantify the success of the channel at preserving the entanglement by computing the concurrence \cite{HillWootters,Wootters:98}, $\mathcal{C}(\rho) = \max(0,\lambda_1 - \lambda_2 - \lambda_3 - \lambda_4)$ where $\lambda_i$ are the eigenvalues in decreasing order of $\sqrt{\sqrt{\rho}Y \rho^* Y \sqrt{\rho}}$, where $Y := \sigma_y \otimes \sigma_y$.

\begin{figure}[h]
\begin{centering}
\includegraphics[scale=0.4]{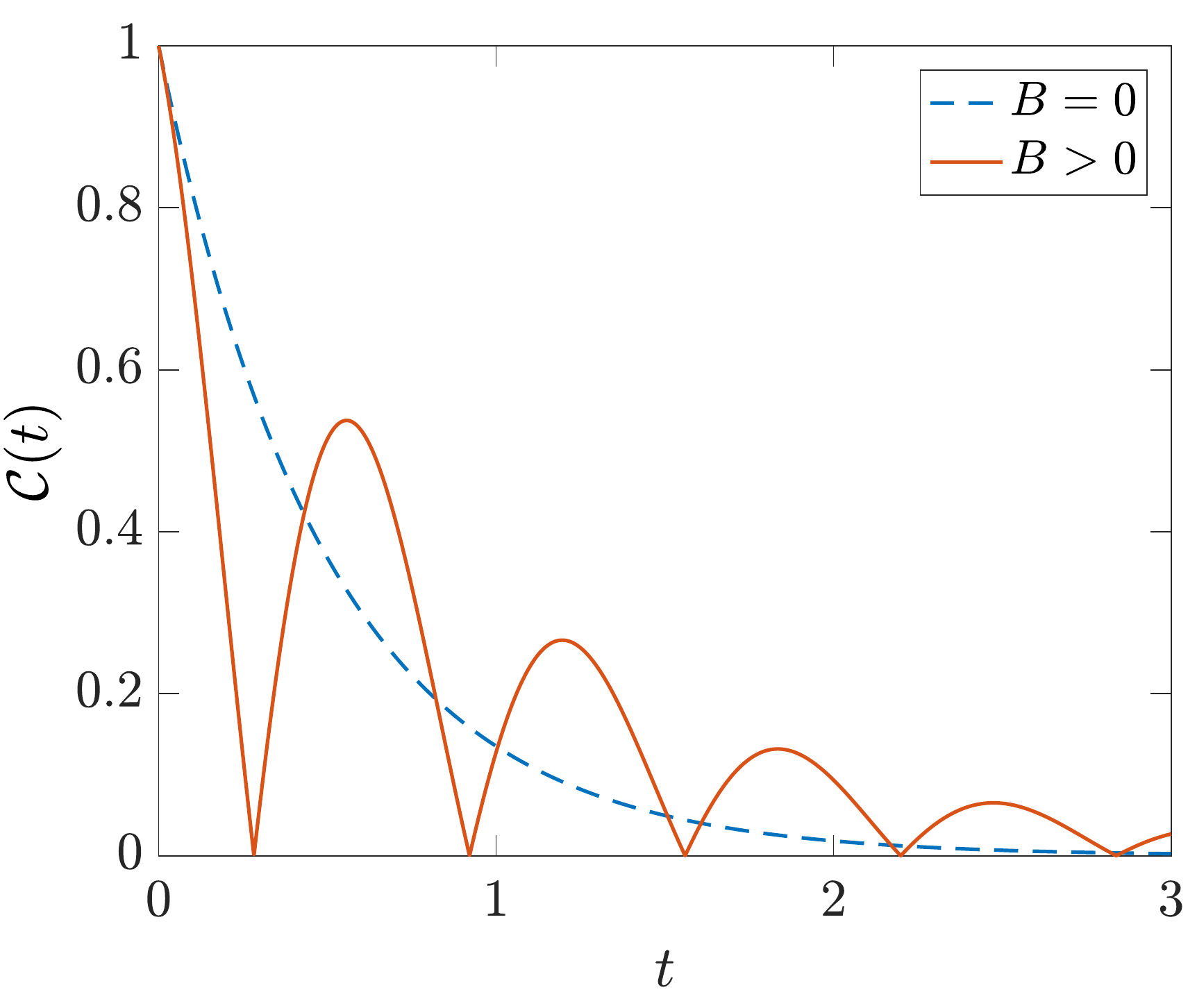}
\end{centering}
\vspace{-4mm}
\caption{Concurrence as a function of time, for an initially maximally entangled pure 2-qubit state, $|\psi\rangle = \frac{1}{\sqrt{2}}(|00\rangle + |11\rangle)$. Here the dephasing is in the $z$-direction acting on one of the qubits. One can show $\mathcal{C}(t)=|\Lambda(t)|$. We plot for both the background channel (blue/dash), and for the channel assisted by the generalized-Markovian noise (red/solid). We see the peaks for the assisted channel surpass the background channel. Here, $\gamma=1$, $B = 5$, $\tau_{k} = 5$. Time is measured in units of $1/ \gamma$.}
\label{concurrence-fig}
\end{figure}

From Fig.~\ref{concurrence-fig}, we again see that, for certain time intervals,  the entanglement of the channel with the added non-trivial memory kernel outperforms the background channel. As before, we find that we can double the channel length, whilst still achieving the same level of entanglement. 

At this point, before looking to more examples, we briefly discuss, in the context
of this example, the physical mechanism
which allows this type of generalized-Markovian noise to protect our system (on some time-scales).

Recall from Eq.~(\ref{ss_proj}) that the `infinite time' state $P_0 \rho$ ($\forall \rho$) does not
evolve (hence decohere)
under action of $\cl{L}_0$ (since $\cl{L}_0P_0 = 0$).
Moreover, from Eq.~(\ref{dephasing}) we can see that 
\begin{equation}
P_0 X = \frac{1}{2}(X + A_k X A_k).
\end{equation}

When we include generalized-Markovian noise in our system,
the dynamics now governed by Eq.~(\ref{newMap}), 
in fact periodically generates such a
`protected' state, $\Phi(t_n)\rho = P_0\rho$,
where (finite) $t_n$ is such that $p(t_n)=1/2$, or equivalently $\Lambda(t_n) = 0$ (i.e. $\cos({\omega t_n + \phi})=0$).
In other words, given some arbitrary initial state $\rho_0$, the time evolved state
$\rho(t) = \Phi(t) \rho_0$, is such that $\mathcal{L}_0\rho(t_n) = 0$.

This shows the system is periodically driven through the steady state of $\cl{L}_0$. At, and close to these times, the Markov part of the dynamics ($\cl{L}_0$ alone) has no, or little, effect. In particular, at these times, the system is essentially decoupled from the environmental noise,
which allows the system to exhibit a lower leading decay rate as compared to a purely Markov evolution, which is subject to the full effects of the decoherence 
induced by $\cl{L}_0$ for all finite $t$.

\subsection{Modulated decay noise}
In this section we briefly study another type of generalized-Markovian process for illustrative purposes, where the long-time decay of the kernel has an additional modulation (i.e., we generalize the previous example). In other words we take $k(t-t') = B^2 e^{-|t-t'|/\tau_k} \cos (\nu(t-t'))$ with Laplace transform given by

\begin{equation}
\tilde{k}(s) = B^2\frac{s + \tau_k^{-1}}{(s+\tau_k^{-1})^2 + \nu^2}.
\end{equation}
 
In order to find the partial fraction decomposition of Eq.~(\ref{eq:PFD}) we need to find the roots of a third order polynomial. 
One can show (see Appendix \ref{sec:cosine}) that taking 
\be
B^2 = \frac{2}{9}(2\gamma - 1/\tau_k)^2 + 2 \nu^2,
\ee
these three roots are given by $s^* = -\tau^{-1}, -\tau^{-1} \pm i \omega$, where 
\begin{eqnarray}
&& \tau^{-1} = \frac{1}{3}(\tau_0^{-1} + 2 \tau_k^{-1}),\\
&& \omega = \sqrt{3 \nu^2 - \frac{1}{9}(2 \gamma - \tau_k^{-1})^2}.
\end{eqnarray}

Note, one is not restricted to taking this choice of $B$, however it is convenient to work with as the decay rate associated with each root is identical ($\tau^{-1}$).

In fact, we have (see Appendix \ref{sec:cosine})
\be
\Lambda(t)=e^{- t/\tau }\left[ c_0 + \frac{1-c_0}{\cos \phi}\cos (\omega t + \phi)\right],
\ee
with
\be
c_0 = \frac{1}{9\omega^2}\left((2\gamma - \tau_k^{-1})^2 + 9\nu^2\right),
\ee
and
\be
\cos \phi = \frac{1-c_0}{\sqrt{(1-c_0)^2 + \frac{4}{9\omega^2} (2 \gamma - \tau_k^{-1})^2}}.
\ee
Since the rate of decay is otherwise given by $2\gamma$ (under $\cl{L}_0$ alone), assuming parameters are chosen so that $\omega \in \mathbb{R}$, the rate of decay can be reduced by up to a factor which approaches  3 in the limit $\tau_k\to \infty$. In fact after evolving the combined channel for times $t_n = 2 \pi n/\omega$ ($n=0,1,2,\dots$), the system is exactly as it would be under evolution of $\cl{L}_0$ alone, with $\gamma \rightarrow \frac{1}{3}(\gamma + \tau_k^{-1})$ (see Appendix \ref{sec:cosine}). We illustrate this in Fig.~\ref{cosine-fig}, where we plot the minimum channel fidelity against time.
We have also numerically verified that the generated dynamics are completely positive for our parameter choices (see Appendix \ref{sec:cosine}).

\begin{figure}
\includegraphics[scale=0.4]{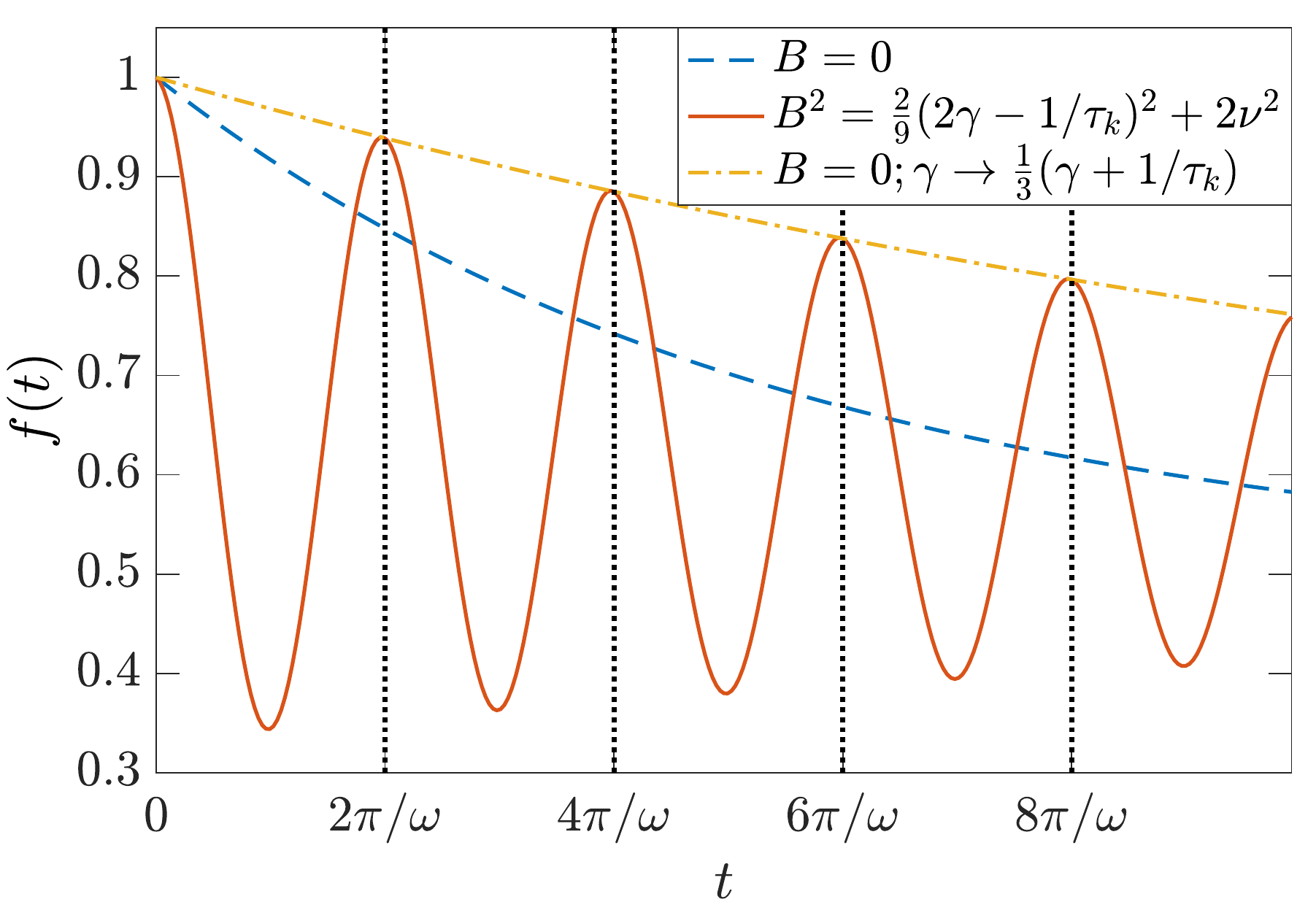}
\caption{Minimum channel fidelity against time for cosine memory kernel. We plot for the background channel (blue/dash, $B=0$), the combined channel (red/solid, $B>0$), and for the background channel with $\gamma$ replaced by $\frac{1}{3}(\gamma+1/\tau_k)$ (yellow/dot-dash). Here $\gamma=0.5,\nu=10,\tau_k=25$.
For this parameter choice, the peaks occur approximately at the times $2\pi n/\omega$. We numerically verify the generated map is CP for all $t\ge 0$ (including the case when we set $\gamma=0$). Time is measured in units of $1/\gamma$.}
\label{cosine-fig}
\end{figure}

\section{Example: Thermal Qubit \label{sec:thermal}}

We provide a further example of our scheme, where the Lindblad operators
defining the Markovian `background' noise are not self-adjoint. 
In the interest of providing an example which is potentially experimentally
verifiable, we restrict our generalized-Markovian noise to the dephasing type (i.e.~with self-adjoint Lindblad operators, see Sect.~\ref{sec:appendix_ME}). 
Note however this makes the analytic solution more complicated (since
$\cl{L}_0$ and $\cl{L}_1$ have different spectra), and therefore we just provide numerics here.
If one does not make such restrictions, then a similar analysis as in the previous
examples can be carried out.

Explicitly, we consider an exponential memory kernel $k(t) = e^{-t/\tau_k}$,
with and  $\cl{L}_0 = \cl{L}_- + \cl{L}_+$, and $\cl{L}_1=\cl{L}_z$, which act as
\begin{eqnarray}
& \cl{L}_\pm X=\gamma_\pm \left( \sigma_\pm X \sigma_{\mp} - \frac{1}{2}\{\sigma_{\mp} \sigma_\pm, X\} \right) \\
& \cl{L}_z X = \gamma_z \left( \sigma_3 X \sigma_3 - X\right)
\end{eqnarray}
where 
$\sigma_\pm := \frac{1}{2}(\sigma_1 \pm i \sigma_2)$. 

Note that evolution under $\cl{L}_0$ alone (i.e. $\gamma_z=0$) indeed generates a
(unique) thermal (Gibbs) state (in the infinite time limit) $\rho_G = \frac{1}{Z}e^{-\beta H}$, at inverse temperature $\beta$ for Hamiltonian $H = g \sigma_3$, where we identify $\gamma_+ /\gamma_- = e^{- \beta g}$ (with normalization $Z = \text{Tr}[e^{-\beta H}]$). 
We also provide a derivation of this in Appendix \ref{app:thermal}.
Also note that $\cl{L}_z \rho_G = 0$ (thermal steady state is a steady state of $\cl{L}_z$).

We study this system numerically, solving it by taking the Laplace transformation (see Eq.~(\ref{general-laplace})).
Note, we check the resulting map $\Phi_t$ is a genuine quantum map for our parameter choices by checking positivity of the Choi matrix \cite{wolf_quantum_2012} \footnote{Choi matrix for map $\Phi$: \unexpanded{$C = \sum_{i,j} |i\rangle \langle j | \otimes \Phi (|i\rangle \langle j|)$}.}.

\begin{figure}
\includegraphics[scale=0.4]{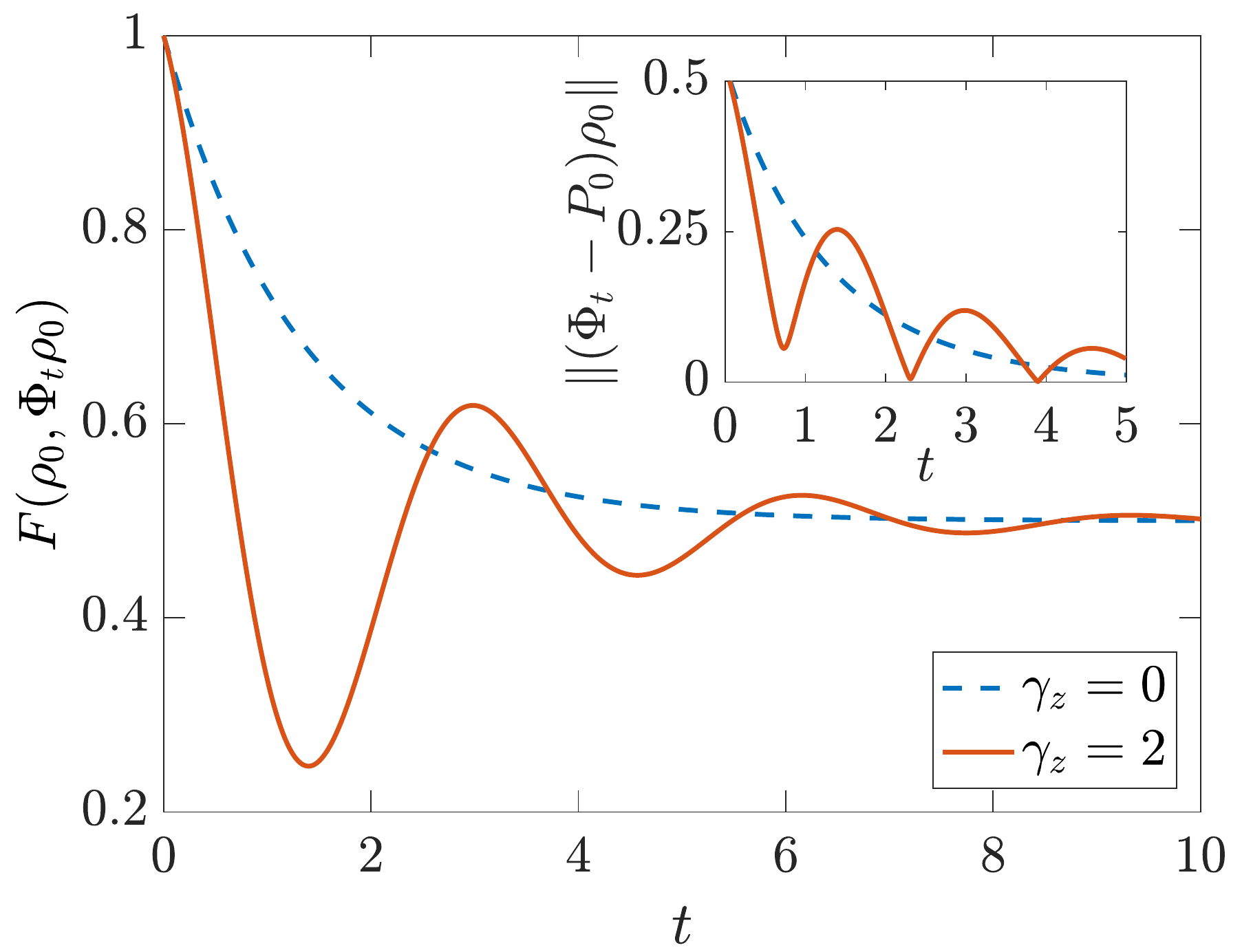}
\caption{Fidelity overlap with initial pure state $\rho_0 = |\psi_0\rangle \langle \psi_0 |$ as a function of time for thermal qubit,
with (red/solid, $\gamma_z =2$) and without (blue/dash, $\gamma_z=0$) added protection by generalized Markovian noise. 
Parameters: $\gamma_- = 1,\gamma_+=1/2,\tau_k=5$.
Initial state: $|\psi_0\rangle = \frac{1}{\sqrt{2}}(|0\rangle + |1\rangle)$. We numerically verify the dynamics are CP (using the Choi matrix).
Inset: Distance of the state at time $t$  to the corresponding steady state of $\cl{L}_0$.
We see periodically the system passes close to the steady state (when $\gamma_z\neq 0$).
When the distance $\|(\Phi_t - P_0)\rho_0\| \approx 0$, the system is essentially decoupled from the thermal noise.
We use the maximum singular value norm.
Time is measured in units of $1/\gamma_-$.
}
\label{thermal-fig}
\end{figure}

We consider the time evolution under $\Phi_t$ of an initial maximal superposition state,
and again find that one can reduce the leading decay rate, e.g., see Fig.~\ref{thermal-fig},
which shows revivals in the fidelity surpassing the background channel.
Moreover, in light of our discussion above, we plot in the inset the distance of
the state at time $t$ under the full evolution (i.e. with the added generalized-Markovian noise) to the corresponding (unique) steady state of the
background dynamics $\cl{L}_0 = \cl{L}_- + \cl{L}_+$, as defined by the projector $P_0:=\lim_{t \rightarrow \infty}e^{t \cl{L}_0}$ (see Eq.~\ref{ss_proj}).
Since the system periodically passes close to the steady state of the background dynamics,
periodically the system is effectively decoupled from this thermal noise.

\section{Discussion}

Researchers into quantum information and open quantum systems are realizing that in some situations, noise can in fact be used to aid in information processing tasks.
 In this work, we have introduced a technique whereby a generalized type of Markovian quantum process can be used to aid in the preservation of quantum information.

In particular, we show that upon adding generalized-Markovian noise on top of an assumed background Markovian dynamics, the rate at which the system decays can in fact be reduced.
The mechanism behind this completely relies on the appearance of the non-trivial memory kernel describing the generalized-Markovian dynamics. One possible way of engineering such dynamics is by introducing a Hamiltonian coupled to a classical stochastic field whose correlation is given by the memory kernel.

We explain this method by considering a Pauli channel, which we analytically solve. We show how an exponential memory kernel can be used to effectively double the length of the channel, whilst still preserving the same threshold for errors, while a cosine-type of kernel  has even greater error suppressing capabilities. Moreover, we find similar results for a qubit in a thermal environment.

We discussed a possible physical mechanism governing these dynamics whereby the system is
periodically driven to (or close to) a steady state of the background dynamics ($\cl{L}_0$),
at which times, the state is essentially decoupled from the background noise.

Remarkably, we have found that the act of adding a certain class of noise to an already dissipative system, can in fact result in less decoherence. This particular technique opens new avenues of study into both dissipation as a resource, and into open systems in general; in particular, at the interface of Markov and non-Markov dynamics, of which there are still many unanswered questions.

\section{Acknowledgments}
The research is based upon work partially supported by the Office of
the Director of National Intelligence (ODNI), Intelligence Advanced
Research Projects Activity (IARPA), via the U.S. Army Research Office
contract W911NF-17-C-0050. The views and conclusions contained herein are
those of the authors and should not be interpreted as necessarily
representing the official policies or endorsements, either expressed or
implied, of the ODNI, IARPA, or the U.S. Government. The U.S. Government
is authorized to reproduce and distribute reprints for Governmental
purposes notwithstanding any copyright annotation thereon.
This work was also partially supported by the ARO MURI grant W911NF-11-1-0268.

\bibliography{refs}

\begin{thebibliography}{41}%
\makeatletter
\providecommand \@ifxundefined [1]{%
 \@ifx{#1\undefined}
}%
\providecommand \@ifnum [1]{%
 \ifnum #1\expandafter \@firstoftwo
 \else \expandafter \@secondoftwo
 \fi
}%
\providecommand \@ifx [1]{%
 \ifx #1\expandafter \@firstoftwo
 \else \expandafter \@secondoftwo
 \fi
}%
\providecommand \natexlab [1]{#1}%
\providecommand \enquote  [1]{``#1''}%
\providecommand \bibnamefont  [1]{#1}%
\providecommand \bibfnamefont [1]{#1}%
\providecommand \citenamefont [1]{#1}%
\providecommand \href@noop [0]{\@secondoftwo}%
\providecommand \href [0]{\begingroup \@sanitize@url \@href}%
\providecommand \@href[1]{\@@startlink{#1}\@@href}%
\providecommand \@@href[1]{\endgroup#1\@@endlink}%
\providecommand \@sanitize@url [0]{\catcode `\\12\catcode `\$12\catcode
  `\&12\catcode `\#12\catcode `\^12\catcode `\_12\catcode `\%12\relax}%
\providecommand \@@startlink[1]{}%
\providecommand \@@endlink[0]{}%
\providecommand \url  [0]{\begingroup\@sanitize@url \@url }%
\providecommand \@url [1]{\endgroup\@href {#1}{\urlprefix }}%
\providecommand \urlprefix  [0]{URL }%
\providecommand \Eprint [0]{\href }%
\providecommand \doibase [0]{http://dx.doi.org/}%
\providecommand \selectlanguage [0]{\@gobble}%
\providecommand \bibinfo  [0]{\@secondoftwo}%
\providecommand \bibfield  [0]{\@secondoftwo}%
\providecommand \translation [1]{[#1]}%
\providecommand \BibitemOpen [0]{}%
\providecommand \bibitemStop [0]{}%
\providecommand \bibitemNoStop [0]{.\EOS\space}%
\providecommand \EOS [0]{\spacefactor3000\relax}%
\providecommand \BibitemShut  [1]{\csname bibitem#1\endcsname}%
\let\auto@bib@innerbib\@empty
\bibitem [{\citenamefont {Landauer}(1995)}]{Landauer:95}%
  \BibitemOpen
  \bibfield  {author} {\bibinfo {author} {\bibfnamefont {R.}~\bibnamefont
  {Landauer}},\ }\href {\doibase 10.1098/rsta.1995.0106} {\bibfield  {journal}
  {\bibinfo  {journal} {Philosophical Transactions of the Royal Society of
  London A: Mathematical, Physical and Engineering Sciences}\ }\textbf
  {\bibinfo {volume} {353}},\ \bibinfo {pages} {367} (\bibinfo {year}
  {1995})}\BibitemShut {NoStop}%
\bibitem [{\citenamefont {Unruh}(1995)}]{Unruh:1995fk}%
  \BibitemOpen
  \bibfield  {author} {\bibinfo {author} {\bibfnamefont {W.~G.}\ \bibnamefont
  {Unruh}},\ }\href {\doibase 10.1103/PhysRevA.51.992} {\bibfield  {journal}
  {\bibinfo  {journal} {{Phys. Rev. A}}\ }\textbf {\bibinfo {volume} {51}},\
  \bibinfo {pages} {992} (\bibinfo {year} {1995})}\BibitemShut {NoStop}%
\bibitem [{\citenamefont {Lidar}\ and\ \citenamefont
  {Brun}(2013)}]{Lidar-Brun:book}%
  \BibitemOpen
  \bibinfo {editor} {\bibfnamefont {D.}~\bibnamefont {Lidar}}\ and\ \bibinfo
  {editor} {\bibfnamefont {T.}~\bibnamefont {Brun}},\ eds.,\ \href
  {http://www.cambridge.org/9780521897877} {\emph {\bibinfo {title} {Quantum
  Error Correction}}}\ (\bibinfo  {publisher} {Cambridge University Press},\
  \bibinfo {address} {{Cambridge, UK}},\ \bibinfo {year} {2013})\BibitemShut
  {NoStop}%
\bibitem [{\citenamefont {Kraus}\ \emph {et~al.}(2008)\citenamefont {Kraus},
  \citenamefont {B{\"u}chler}, \citenamefont {Diehl}, \citenamefont {Kantian},
  \citenamefont {Micheli},\ and\ \citenamefont {Zoller}}]{Kraus-prep}%
  \BibitemOpen
  \bibfield  {author} {\bibinfo {author} {\bibfnamefont {B.}~\bibnamefont
  {Kraus}}, \bibinfo {author} {\bibfnamefont {H.~P.}\ \bibnamefont
  {B{\"u}chler}}, \bibinfo {author} {\bibfnamefont {S.}~\bibnamefont {Diehl}},
  \bibinfo {author} {\bibfnamefont {A.}~\bibnamefont {Kantian}}, \bibinfo
  {author} {\bibfnamefont {A.}~\bibnamefont {Micheli}}, \ and\ \bibinfo
  {author} {\bibfnamefont {P.}~\bibnamefont {Zoller}},\ }\href
  {http://link.aps.org/doi/10.1103/PhysRevA.78.042307} {\bibfield  {journal}
  {\bibinfo  {journal} {Physical Review A}\ }\textbf {\bibinfo {volume} {78}},\
  \bibinfo {pages} {042307} (\bibinfo {year} {2008})}\BibitemShut {NoStop}%
\bibitem [{\citenamefont {Verstraete}\ \emph {et~al.}(2009)\citenamefont
  {Verstraete}, \citenamefont {Wolf},\ and\ \citenamefont
  {Ignacio~Cirac}}]{verstraete2009quantum}%
  \BibitemOpen
  \bibfield  {author} {\bibinfo {author} {\bibfnamefont {F.}~\bibnamefont
  {Verstraete}}, \bibinfo {author} {\bibfnamefont {M.~M.}\ \bibnamefont
  {Wolf}}, \ and\ \bibinfo {author} {\bibfnamefont {J.}~\bibnamefont
  {Ignacio~Cirac}},\ }\href {http://dx.doi.org/10.1038/nphys1342} {\bibfield
  {journal} {\bibinfo  {journal} {Nat Phys}\ }\textbf {\bibinfo {volume} {5}},\
  \bibinfo {pages} {633} (\bibinfo {year} {2009})}\BibitemShut {NoStop}%
\bibitem [{\citenamefont {Barreiro}\ \emph {et~al.}(2011)\citenamefont
  {Barreiro}, \citenamefont {Muller}, \citenamefont {Schindler}, \citenamefont
  {Nigg}, \citenamefont {Monz}, \citenamefont {Chwalla}, \citenamefont
  {Hennrich}, \citenamefont {Roos}, \citenamefont {Zoller},\ and\ \citenamefont
  {Blatt}}]{barreiro2011open}%
  \BibitemOpen
  \bibfield  {author} {\bibinfo {author} {\bibfnamefont {J.~T.}\ \bibnamefont
  {Barreiro}}, \bibinfo {author} {\bibfnamefont {M.}~\bibnamefont {Muller}},
  \bibinfo {author} {\bibfnamefont {P.}~\bibnamefont {Schindler}}, \bibinfo
  {author} {\bibfnamefont {D.}~\bibnamefont {Nigg}}, \bibinfo {author}
  {\bibfnamefont {T.}~\bibnamefont {Monz}}, \bibinfo {author} {\bibfnamefont
  {M.}~\bibnamefont {Chwalla}}, \bibinfo {author} {\bibfnamefont
  {M.}~\bibnamefont {Hennrich}}, \bibinfo {author} {\bibfnamefont {C.~F.}\
  \bibnamefont {Roos}}, \bibinfo {author} {\bibfnamefont {P.}~\bibnamefont
  {Zoller}}, \ and\ \bibinfo {author} {\bibfnamefont {R.}~\bibnamefont
  {Blatt}},\ }\href {http://dx.doi.org/10.1038/nature09801} {\bibfield
  {journal} {\bibinfo  {journal} {Nature}\ }\textbf {\bibinfo {volume} {470}},\
  \bibinfo {pages} {486} (\bibinfo {year} {2011})}\BibitemShut {NoStop}%
\bibitem [{\citenamefont {Wang}\ \emph {et~al.}(2011)\citenamefont {Wang},
  \citenamefont {Ashhab},\ and\ \citenamefont {Nori}}]{PhysRevA.83.062317}%
  \BibitemOpen
  \bibfield  {author} {\bibinfo {author} {\bibfnamefont {H.}~\bibnamefont
  {Wang}}, \bibinfo {author} {\bibfnamefont {S.}~\bibnamefont {Ashhab}}, \ and\
  \bibinfo {author} {\bibfnamefont {F.}~\bibnamefont {Nori}},\ }\href {\doibase
  10.1103/PhysRevA.83.062317} {\bibfield  {journal} {\bibinfo  {journal} {Phys.
  Rev. A}\ }\textbf {\bibinfo {volume} {83}},\ \bibinfo {pages} {062317}
  (\bibinfo {year} {2011})}\BibitemShut {NoStop}%
\bibitem [{\citenamefont {Barthel}\ and\ \citenamefont
  {Kliesch}(2012)}]{markov_simulation_barthel}%
  \BibitemOpen
  \bibfield  {author} {\bibinfo {author} {\bibfnamefont {T.}~\bibnamefont
  {Barthel}}\ and\ \bibinfo {author} {\bibfnamefont {M.}~\bibnamefont
  {Kliesch}},\ }\href {\doibase 10.1103/PhysRevLett.108.230504} {\bibfield
  {journal} {\bibinfo  {journal} {Phys. Rev. Lett.}\ }\textbf {\bibinfo
  {volume} {108}},\ \bibinfo {pages} {230504} (\bibinfo {year}
  {2012})}\BibitemShut {NoStop}%
\bibitem [{\citenamefont {Zanardi}\ and\ \citenamefont
  {Campos~Venuti}(2014)}]{zanardi-dissipation-2014}%
  \BibitemOpen
  \bibfield  {author} {\bibinfo {author} {\bibfnamefont {P.}~\bibnamefont
  {Zanardi}}\ and\ \bibinfo {author} {\bibfnamefont {L.}~\bibnamefont
  {Campos~Venuti}},\ }\href
  {http://journals.aps.org/prl/abstract/10.1103/PhysRevLett.113.240406}
  {\bibfield  {journal} {\bibinfo  {journal} {{Phys.~Rev.~Lett.}}\ }\textbf
  {\bibinfo {volume} {113}},\ \bibinfo {pages} {240406} (\bibinfo {year}
  {2014})}\BibitemShut {NoStop}%
\bibitem [{\citenamefont {Zanardi}\ \emph {et~al.}(2016)\citenamefont
  {Zanardi}, \citenamefont {Marshall},\ and\ \citenamefont
  {Campos~Venuti}}]{dissi_2nd_order:Zanardi2016}%
  \BibitemOpen
  \bibfield  {author} {\bibinfo {author} {\bibfnamefont {P.}~\bibnamefont
  {Zanardi}}, \bibinfo {author} {\bibfnamefont {J.}~\bibnamefont {Marshall}}, \
  and\ \bibinfo {author} {\bibfnamefont {L.}~\bibnamefont {Campos~Venuti}},\
  }\href {\doibase 10.1103/PhysRevA.93.022312} {\bibfield  {journal} {\bibinfo
  {journal} {Phys. Rev. A}\ }\textbf {\bibinfo {volume} {93}},\ \bibinfo
  {pages} {022312} (\bibinfo {year} {2016})}\BibitemShut {NoStop}%
\bibitem [{\citenamefont {Marshall}\ \emph {et~al.}(2016)\citenamefont
  {Marshall}, \citenamefont {Campos~Venuti},\ and\ \citenamefont
  {Zanardi}}]{DGM}%
  \BibitemOpen
  \bibfield  {author} {\bibinfo {author} {\bibfnamefont {J.}~\bibnamefont
  {Marshall}}, \bibinfo {author} {\bibfnamefont {L.}~\bibnamefont
  {Campos~Venuti}}, \ and\ \bibinfo {author} {\bibfnamefont {P.}~\bibnamefont
  {Zanardi}},\ }\href {\doibase 10.1103/PhysRevA.94.052339} {\bibfield
  {journal} {\bibinfo  {journal} {Phys. Rev. A}\ }\textbf {\bibinfo {volume}
  {94}},\ \bibinfo {pages} {052339} (\bibinfo {year} {2016})}\BibitemShut
  {NoStop}%
\bibitem [{\citenamefont {Chiuri}\ \emph {et~al.}(2012)\citenamefont {Chiuri},
  \citenamefont {Greganti}, \citenamefont {Mazzola}, \citenamefont
  {Paternostro},\ and\ \citenamefont {Mataloni}}]{NM_simulation_12}%
  \BibitemOpen
  \bibfield  {author} {\bibinfo {author} {\bibfnamefont {A.}~\bibnamefont
  {Chiuri}}, \bibinfo {author} {\bibfnamefont {C.}~\bibnamefont {Greganti}},
  \bibinfo {author} {\bibfnamefont {L.}~\bibnamefont {Mazzola}}, \bibinfo
  {author} {\bibfnamefont {M.}~\bibnamefont {Paternostro}}, \ and\ \bibinfo
  {author} {\bibfnamefont {P.}~\bibnamefont {Mataloni}},\ }\href
  {http://dx.doi.org/10.1038/srep00968} {\bibfield  {journal} {\bibinfo
  {journal} {Scientific Reports}\ }\textbf {\bibinfo {volume} {2}},\ \bibinfo
  {pages} {968} (\bibinfo {year} {2012})}\BibitemShut {NoStop}%
\bibitem [{\citenamefont {C\'ardenas}\ \emph {et~al.}(2015)\citenamefont
  {C\'ardenas}, \citenamefont {Paternostro},\ and\ \citenamefont
  {Semi\~ao}}]{NM_simulation_15}%
  \BibitemOpen
  \bibfield  {author} {\bibinfo {author} {\bibfnamefont {P.~C.}\ \bibnamefont
  {C\'ardenas}}, \bibinfo {author} {\bibfnamefont {M.}~\bibnamefont
  {Paternostro}}, \ and\ \bibinfo {author} {\bibfnamefont {F.~L.}\ \bibnamefont
  {Semi\~ao}},\ }\href {\doibase 10.1103/PhysRevA.91.022122} {\bibfield
  {journal} {\bibinfo  {journal} {Phys. Rev. A}\ }\textbf {\bibinfo {volume}
  {91}},\ \bibinfo {pages} {022122} (\bibinfo {year} {2015})}\BibitemShut
  {NoStop}%
\bibitem [{\citenamefont {Brito}\ and\ \citenamefont
  {Werlang}(2015)}]{markovianity_tunable}%
  \BibitemOpen
  \bibfield  {author} {\bibinfo {author} {\bibfnamefont {F.}~\bibnamefont
  {Brito}}\ and\ \bibinfo {author} {\bibfnamefont {T.}~\bibnamefont
  {Werlang}},\ }\href {http://stacks.iop.org/1367-2630/17/i=7/a=072001}
  {\bibfield  {journal} {\bibinfo  {journal} {New Journal of Physics}\ }\textbf
  {\bibinfo {volume} {17}},\ \bibinfo {pages} {072001} (\bibinfo {year}
  {2015})}\BibitemShut {NoStop}%
\bibitem [{\citenamefont {Sweke}\ \emph {et~al.}(2016)\citenamefont {Sweke},
  \citenamefont {Sanz}, \citenamefont {Sinayskiy}, \citenamefont
  {Petruccione},\ and\ \citenamefont {Solano}}]{NM_sweke}%
  \BibitemOpen
  \bibfield  {author} {\bibinfo {author} {\bibfnamefont {R.}~\bibnamefont
  {Sweke}}, \bibinfo {author} {\bibfnamefont {M.}~\bibnamefont {Sanz}},
  \bibinfo {author} {\bibfnamefont {I.}~\bibnamefont {Sinayskiy}}, \bibinfo
  {author} {\bibfnamefont {F.}~\bibnamefont {Petruccione}}, \ and\ \bibinfo
  {author} {\bibfnamefont {E.}~\bibnamefont {Solano}},\ }\href {\doibase
  10.1103/PhysRevA.94.022317} {\bibfield  {journal} {\bibinfo  {journal} {Phys.
  Rev. A}\ }\textbf {\bibinfo {volume} {94}},\ \bibinfo {pages} {022317}
  (\bibinfo {year} {2016})}\BibitemShut {NoStop}%
\bibitem [{\citenamefont {Di~Candia}\ \emph {et~al.}(2015)\citenamefont
  {Di~Candia}, \citenamefont {Pedernales}, \citenamefont {del Campo},
  \citenamefont {Solano},\ and\ \citenamefont {Casanova}}]{simulation_disi15}%
  \BibitemOpen
  \bibfield  {author} {\bibinfo {author} {\bibfnamefont {R.}~\bibnamefont
  {Di~Candia}}, \bibinfo {author} {\bibfnamefont {J.~S.}\ \bibnamefont
  {Pedernales}}, \bibinfo {author} {\bibfnamefont {A.}~\bibnamefont {del
  Campo}}, \bibinfo {author} {\bibfnamefont {E.}~\bibnamefont {Solano}}, \ and\
  \bibinfo {author} {\bibfnamefont {J.}~\bibnamefont {Casanova}},\ }\href
  {http://dx.doi.org/10.1038/srep09981} {\bibfield  {journal} {\bibinfo
  {journal} {Scientific Reports}\ }\textbf {\bibinfo {volume} {5}},\ \bibinfo
  {pages} {9981} (\bibinfo {year} {2015})}\BibitemShut {NoStop}%
\bibitem [{\citenamefont {Poyatos}\ \emph {et~al.}(1996)\citenamefont
  {Poyatos}, \citenamefont {Cirac},\ and\ \citenamefont
  {Zoller}}]{res_eng_poyatos}%
  \BibitemOpen
  \bibfield  {author} {\bibinfo {author} {\bibfnamefont {J.~F.}\ \bibnamefont
  {Poyatos}}, \bibinfo {author} {\bibfnamefont {J.~I.}\ \bibnamefont {Cirac}},
  \ and\ \bibinfo {author} {\bibfnamefont {P.}~\bibnamefont {Zoller}},\ }\href
  {\doibase 10.1103/PhysRevLett.77.4728} {\bibfield  {journal} {\bibinfo
  {journal} {Phys. Rev. Lett.}\ }\textbf {\bibinfo {volume} {77}},\ \bibinfo
  {pages} {4728} (\bibinfo {year} {1996})}\BibitemShut {NoStop}%
\bibitem [{\citenamefont {Carvalho}\ \emph {et~al.}(2001)\citenamefont
  {Carvalho}, \citenamefont {Milman}, \citenamefont {de~Matos~Filho},\ and\
  \citenamefont {Davidovich}}]{PhysRevLett.86.4988}%
  \BibitemOpen
  \bibfield  {author} {\bibinfo {author} {\bibfnamefont {A.~R.~R.}\
  \bibnamefont {Carvalho}}, \bibinfo {author} {\bibfnamefont {P.}~\bibnamefont
  {Milman}}, \bibinfo {author} {\bibfnamefont {R.~L.}\ \bibnamefont
  {de~Matos~Filho}}, \ and\ \bibinfo {author} {\bibfnamefont {L.}~\bibnamefont
  {Davidovich}},\ }\href {\doibase 10.1103/PhysRevLett.86.4988} {\bibfield
  {journal} {\bibinfo  {journal} {Phys. Rev. Lett.}\ }\textbf {\bibinfo
  {volume} {86}},\ \bibinfo {pages} {4988} (\bibinfo {year}
  {2001})}\BibitemShut {NoStop}%
\bibitem [{\citenamefont {Bellomo}\ \emph {et~al.}(2008)\citenamefont
  {Bellomo}, \citenamefont {Lo~Franco}, \citenamefont {Maniscalco},\ and\
  \citenamefont {Compagno}}]{Bellomo_structured_env}%
  \BibitemOpen
  \bibfield  {author} {\bibinfo {author} {\bibfnamefont {B.}~\bibnamefont
  {Bellomo}}, \bibinfo {author} {\bibfnamefont {R.}~\bibnamefont {Lo~Franco}},
  \bibinfo {author} {\bibfnamefont {S.}~\bibnamefont {Maniscalco}}, \ and\
  \bibinfo {author} {\bibfnamefont {G.}~\bibnamefont {Compagno}},\ }\href
  {\doibase 10.1103/PhysRevA.78.060302} {\bibfield  {journal} {\bibinfo
  {journal} {Phys. Rev. A}\ }\textbf {\bibinfo {volume} {78}},\ \bibinfo
  {pages} {060302} (\bibinfo {year} {2008})}\BibitemShut {NoStop}%
\bibitem [{\citenamefont {Tan}\ \emph {et~al.}(2010)\citenamefont {Tan},
  \citenamefont {Kyaw},\ and\ \citenamefont {Yeo}}]{PhysRevA.81.062119}%
  \BibitemOpen
  \bibfield  {author} {\bibinfo {author} {\bibfnamefont {J.}~\bibnamefont
  {Tan}}, \bibinfo {author} {\bibfnamefont {T.~H.}\ \bibnamefont {Kyaw}}, \
  and\ \bibinfo {author} {\bibfnamefont {Y.}~\bibnamefont {Yeo}},\ }\href
  {\doibase 10.1103/PhysRevA.81.062119} {\bibfield  {journal} {\bibinfo
  {journal} {Phys. Rev. A}\ }\textbf {\bibinfo {volume} {81}},\ \bibinfo
  {pages} {062119} (\bibinfo {year} {2010})}\BibitemShut {NoStop}%
\bibitem [{\citenamefont {Schirmer}\ and\ \citenamefont
  {Wang}(2010)}]{PhysRevA.81.062306}%
  \BibitemOpen
  \bibfield  {author} {\bibinfo {author} {\bibfnamefont {S.~G.}\ \bibnamefont
  {Schirmer}}\ and\ \bibinfo {author} {\bibfnamefont {X.}~\bibnamefont
  {Wang}},\ }\href {\doibase 10.1103/PhysRevA.81.062306} {\bibfield  {journal}
  {\bibinfo  {journal} {Phys. Rev. A}\ }\textbf {\bibinfo {volume} {81}},\
  \bibinfo {pages} {062306} (\bibinfo {year} {2010})}\BibitemShut {NoStop}%
\bibitem [{\citenamefont {Tong}\ \emph {et~al.}(2010)\citenamefont {Tong},
  \citenamefont {An}, \citenamefont {Luo},\ and\ \citenamefont
  {Oh}}]{PhysRevA.81.052330}%
  \BibitemOpen
  \bibfield  {author} {\bibinfo {author} {\bibfnamefont {Q.-J.}\ \bibnamefont
  {Tong}}, \bibinfo {author} {\bibfnamefont {J.-H.}\ \bibnamefont {An}},
  \bibinfo {author} {\bibfnamefont {H.-G.}\ \bibnamefont {Luo}}, \ and\
  \bibinfo {author} {\bibfnamefont {C.~H.}\ \bibnamefont {Oh}},\ }\href
  {\doibase 10.1103/PhysRevA.81.052330} {\bibfield  {journal} {\bibinfo
  {journal} {Phys. Rev. A}\ }\textbf {\bibinfo {volume} {81}},\ \bibinfo
  {pages} {052330} (\bibinfo {year} {2010})}\BibitemShut {NoStop}%
\bibitem [{\citenamefont {Man}\ \emph {et~al.}(2015)\citenamefont {Man},
  \citenamefont {Xia},\ and\ \citenamefont {Lo~Franco}}]{man:2015_cavity}%
  \BibitemOpen
  \bibfield  {author} {\bibinfo {author} {\bibfnamefont {Z.-X.}\ \bibnamefont
  {Man}}, \bibinfo {author} {\bibfnamefont {Y.-J.}\ \bibnamefont {Xia}}, \ and\
  \bibinfo {author} {\bibfnamefont {R.}~\bibnamefont {Lo~Franco}},\ }\href
  {http://dx.doi.org/10.1038/srep13843} {\bibfield  {journal} {\bibinfo
  {journal} {Scientific Reports}\ }\textbf {\bibinfo {volume} {5}},\ \bibinfo
  {pages} {13843} (\bibinfo {year} {2015})}\BibitemShut {NoStop}%
\bibitem [{\citenamefont {Shabani}\ and\ \citenamefont
  {Lidar}(2005)}]{ShabaniLidar:05}%
  \BibitemOpen
  \bibfield  {author} {\bibinfo {author} {\bibfnamefont {A.}~\bibnamefont
  {Shabani}}\ and\ \bibinfo {author} {\bibfnamefont {D.~A.}\ \bibnamefont
  {Lidar}},\ }\href {\doibase 10.1103/PhysRevA.71.020101} {\bibfield  {journal}
  {\bibinfo  {journal} {Phys. Rev. A}\ }\textbf {\bibinfo {volume} {71}},\
  \bibinfo {pages} {020101} (\bibinfo {year} {2005})}\BibitemShut {NoStop}%
\bibitem [{\citenamefont {Vacchini}(2016)}]{vacchiniME}%
  \BibitemOpen
  \bibfield  {author} {\bibinfo {author} {\bibfnamefont {B.}~\bibnamefont
  {Vacchini}},\ }\href {\doibase 10.1103/PhysRevLett.117.230401} {\bibfield
  {journal} {\bibinfo  {journal} {Phys. Rev. Lett.}\ }\textbf {\bibinfo
  {volume} {117}},\ \bibinfo {pages} {230401} (\bibinfo {year}
  {2016})}\BibitemShut {NoStop}%
\bibitem [{\citenamefont {Chru\ifmmode \acute{s}\else
  \'{s}\fi{}ci\ifmmode~\acute{n}\else \'{n}\fi{}ski}\ and\ \citenamefont
  {Kossakowski}(2017)}]{semi_markov_pra}%
  \BibitemOpen
  \bibfield  {author} {\bibinfo {author} {\bibfnamefont {D.}~\bibnamefont
  {Chru\ifmmode \acute{s}\else \'{s}\fi{}ci\ifmmode~\acute{n}\else
  \'{n}\fi{}ski}}\ and\ \bibinfo {author} {\bibfnamefont {A.}~\bibnamefont
  {Kossakowski}},\ }\href {\doibase 10.1103/PhysRevA.95.042131} {\bibfield
  {journal} {\bibinfo  {journal} {Phys. Rev. A}\ }\textbf {\bibinfo {volume}
  {95}},\ \bibinfo {pages} {042131} (\bibinfo {year} {2017})}\BibitemShut
  {NoStop}%
\bibitem [{\citenamefont {Lindblad}(1976)}]{Lindblad:76}%
  \BibitemOpen
  \bibfield  {author} {\bibinfo {author} {\bibfnamefont {G.}~\bibnamefont
  {Lindblad}},\ }\href {\doibase 10.1007/BF01608499} {\bibfield  {journal}
  {\bibinfo  {journal} {Comm. Math. Phys.}\ }\textbf {\bibinfo {volume} {48}},\
  \bibinfo {pages} {119} (\bibinfo {year} {1976})}\BibitemShut {NoStop}%
\bibitem [{Note1()}]{Note1}%
  \BibitemOpen
  \bibinfo {note} {$\protect \mathaccentV {dot}05FX := dX/dt$}\BibitemShut
  {NoStop}%
\bibitem [{\citenamefont {{T. Kato}}(1995)}]{Kato:book}%
  \BibitemOpen
  \bibfield  {author} {\bibinfo {author} {\bibnamefont {{T. Kato}}},\
  }\href@noop {} {\emph {\bibinfo {title} {{Perturbation Theory for Linear
  Operators}}}},\ {Classics in Mathematics}\ (\bibinfo  {publisher}
  {{Springer-Verlag}},\ \bibinfo {address} {{Berlin}},\ \bibinfo {year}
  {1995})\BibitemShut {NoStop}%
\bibitem [{\citenamefont {Wolf}()}]{wolf_quantum_2012}%
  \BibitemOpen
  \bibfield  {author} {\bibinfo {author} {\bibfnamefont {M.~M.}\ \bibnamefont
  {Wolf}},\ }\href@noop {} {\enquote {\bibinfo {title} {{Quantum Channels \&
  Operations: Guided Tour}},}\ }\bibinfo {note}
  {\href{https://www-m5.ma.tum.de/foswiki/pub/M5/Allgemeines/MichaelWolf/QChannelLecture.pdf}{Lecture
  notes available online (2012)}}\BibitemShut {NoStop}%
\bibitem [{\citenamefont {Campos~Venuti}\ \emph {et~al.}(2016)\citenamefont
  {Campos~Venuti}, \citenamefont {Albash}, \citenamefont {Lidar},\ and\
  \citenamefont {Zanardi}}]{venuti_adiabaticity_2016}%
  \BibitemOpen
  \bibfield  {author} {\bibinfo {author} {\bibfnamefont {L.}~\bibnamefont
  {Campos~Venuti}}, \bibinfo {author} {\bibfnamefont {T.}~\bibnamefont
  {Albash}}, \bibinfo {author} {\bibfnamefont {D.~A.}\ \bibnamefont {Lidar}}, \
  and\ \bibinfo {author} {\bibfnamefont {P.}~\bibnamefont {Zanardi}},\ }\href
  {http://link.aps.org/doi/10.1103/PhysRevA.93.032118} {\bibfield  {journal}
  {\bibinfo  {journal} {Phys. Rev. A}\ }\textbf {\bibinfo {volume} {93}},\
  \bibinfo {pages} {032118} (\bibinfo {year} {2016})}\BibitemShut {NoStop}%
\bibitem [{\citenamefont {Barnett}\ and\ \citenamefont
  {Stenholm}(2001)}]{hazards}%
  \BibitemOpen
  \bibfield  {author} {\bibinfo {author} {\bibfnamefont {S.~M.}\ \bibnamefont
  {Barnett}}\ and\ \bibinfo {author} {\bibfnamefont {S.}~\bibnamefont
  {Stenholm}},\ }\href {\doibase 10.1103/PhysRevA.64.033808} {\bibfield
  {journal} {\bibinfo  {journal} {Phys. Rev. A}\ }\textbf {\bibinfo {volume}
  {64}},\ \bibinfo {pages} {033808} (\bibinfo {year} {2001})}\BibitemShut
  {NoStop}%
\bibitem [{\citenamefont {Daffer}\ \emph {et~al.}(2004)\citenamefont {Daffer},
  \citenamefont {W\'odkiewicz}, \citenamefont {Cresser},\ and\ \citenamefont
  {McIver}}]{PhysRevA.70.010304}%
  \BibitemOpen
  \bibfield  {author} {\bibinfo {author} {\bibfnamefont {S.}~\bibnamefont
  {Daffer}}, \bibinfo {author} {\bibfnamefont {K.}~\bibnamefont
  {W\'odkiewicz}}, \bibinfo {author} {\bibfnamefont {J.~D.}\ \bibnamefont
  {Cresser}}, \ and\ \bibinfo {author} {\bibfnamefont {J.~K.}\ \bibnamefont
  {McIver}},\ }\href {\doibase 10.1103/PhysRevA.70.010304} {\bibfield
  {journal} {\bibinfo  {journal} {Phys. Rev. A}\ }\textbf {\bibinfo {volume}
  {70}},\ \bibinfo {pages} {010304} (\bibinfo {year} {2004})}\BibitemShut
  {NoStop}%
\bibitem [{Note2()}]{Note2}%
  \BibitemOpen
  \bibinfo {note} {One can think of this as a model for a classically
  correlated noisy channel (if an error occurs, it occurs to all qubits
  simultaneously). See e.g.~Ref.~\cite {dynamicalMemory} for a two qubit
  version.}\BibitemShut {Stop}%
\bibitem [{Note3()}]{Note3}%
  \BibitemOpen
  \bibinfo {note} {We will occasionally make use of the spectral decomposition
  of $\sigma _3 = |1\rangle \langle 1| - |0\rangle \langle 0|$. Throughout,
  when we write $|0\rangle ,|1\rangle $, it is in this
  $z$-eigenbasis.}\BibitemShut {Stop}%
\bibitem [{\citenamefont {{S. Hill and W. K. Wootters}}(1997)}]{HillWootters}%
  \BibitemOpen
  \bibfield  {author} {\bibinfo {author} {\bibnamefont {{S. Hill and W. K.
  Wootters}}},\ }\href {\doibase 10.1103/PhysRevLett.78.5022} {\bibfield
  {journal} {\bibinfo  {journal} {Phys. Rev. Lett.}\ }\textbf {\bibinfo
  {volume} {78}},\ \bibinfo {pages} {5022} (\bibinfo {year}
  {1997})}\BibitemShut {NoStop}%
\bibitem [{\citenamefont {{W.K. Wootters}}(1998)}]{Wootters:98}%
  \BibitemOpen
  \bibfield  {author} {\bibinfo {author} {\bibnamefont {{W.K. Wootters}}},\
  }\href {\doibase 10.1103/PhysRevLett.80.2245} {\bibfield  {journal} {\bibinfo
   {journal} {Phys. Rev. Lett.}\ }\textbf {\bibinfo {volume} {80}},\ \bibinfo
  {pages} {2245} (\bibinfo {year} {1998})}\BibitemShut {NoStop}%
\bibitem [{Note4()}]{Note4}%
  \BibitemOpen
  \bibinfo {note} {Choi matrix for map $\Phi $: $C = \sum _{i,j} |i\rangle
  \langle j | \otimes \Phi (|i\rangle \langle j|)$.}\BibitemShut {Stop}%
\bibitem [{\citenamefont {Addis}\ \emph {et~al.}(2016)\citenamefont {Addis},
  \citenamefont {Karpat}, \citenamefont {Macchiavello},\ and\ \citenamefont
  {Maniscalco}}]{dynamicalMemory}%
  \BibitemOpen
  \bibfield  {author} {\bibinfo {author} {\bibfnamefont {C.}~\bibnamefont
  {Addis}}, \bibinfo {author} {\bibfnamefont {G.}~\bibnamefont {Karpat}},
  \bibinfo {author} {\bibfnamefont {C.}~\bibnamefont {Macchiavello}}, \ and\
  \bibinfo {author} {\bibfnamefont {S.}~\bibnamefont {Maniscalco}},\ }\href
  {\doibase 10.1103/PhysRevA.94.032121} {\bibfield  {journal} {\bibinfo
  {journal} {Phys. Rev. A}\ }\textbf {\bibinfo {volume} {94}},\ \bibinfo
  {pages} {032121} (\bibinfo {year} {2016})}\BibitemShut {NoStop}%
\bibitem [{Note5()}]{Note5}%
  \BibitemOpen
  \bibinfo {note} {This type of process is known as wide-sense
  stationary.}\BibitemShut {Stop}%
\bibitem [{Note6()}]{Note6}%
  \BibitemOpen
  \bibinfo {note} {Note, if in fact $\omega =0$, one can easily see directly
  that $|\Lambda (t)| \le 1$.}\BibitemShut {Stop}%
\end{thebibliography}%

\newpage

\appendix
\section{}

\subsection{Stochastic Hamiltonian derivation of Eq.~(\ref{masterEq}) for self-adjoint Lindblad operators \label{sec:appendix_ME}}

Let us consider adding a stochastic Hamiltonian, $H(t) = B(t) h$, on top of our background dissipative dynamics so that the time evolution is described by
\begin{equation}
\label{stochastic}
\dot \rho (t) = \mathcal{L}_0\rho(t) -i[H(t),\rho(t)]
\end{equation}
where $h=h^\dagger$ is time-independent, and $B(t) \in \mathbb{R}$ is a stochastic variable (we use the convention that $\hbar=1$).
We assume the statistics governing the underlying stochastic process is such that $\langle B(t)\rangle=0$, and $\langle B(t) B(t')\rangle = k(t-t')$ \footnote{This type of process is known as wide-sense stationary.}, where the angle brackets indicate averaging over independent trials.

We will average out the stochastic noise to arrive at a noise-averaged description of the dynamics - we closely follow the derivation in Ref.~\cite{PhysRevA.70.010304}. First note that one can formally solve Eq.~(\ref{stochastic}) as
\begin{equation}
\rho(t) = \rho(0) + \mathcal{L}_0 \int_0^t \rho(t')dt' -i\int_0^t [H(t'),\rho(t')]dt',
\end{equation}
which can be re-inserted into the right hand side of Eq.~(\ref{stochastic}):

\begin{equation}
\begin{split}
\dot \rho(t) = \mathcal{L}_0\rho(t) -iB(t)[h,\rho(0)] -i\mathcal{L}_0 \int_0^t B(t)[h,\rho(t')]dt' \\
- \int_0^t B(t)B(t')[h,[h,\rho(t')]]dt'
\end{split}
\end{equation}

If we assume that the state is sufficiently decorrelated from the random variables [e.g. $\langle B(t)B(t')\rho(t')\rangle \approx \langle B(t)B(t')\rangle \langle \rho(t')\rangle$], performing the averaging as above, we get an equation for the noise-averaged density operator (we drop the angle bracket notation on $\rho$):
\begin{equation}
\dot \rho(t) = \mathcal{L}_0\rho(t)  + \mathcal{L}_1 \int_0^t k(t-t')\rho(t')dt',
\end{equation}
where $\mathcal{L}_1 X = 2 hXh - \{h^2,X\}$. We note that the term $\mathcal{L}_1$ is in Lindblad form, with self-adjoint Lindblad jump operator $h$.

Note that taking a sum $H(t) = \sum B_i(t) h_i$, with $\langle B_i(t) B_j(t') \rangle = \delta_{ij}k(t-t')$ allows one to generate a sum of such Lindbladian generators
(each with self-adjoint Lindblad operators).

\subsection{Derivation of Eqs.~(\ref{spectral}) and (\ref{projectors}) \label{pauli-deriv}}
The easiest way to see the spectral projection for generator Eq.~(\ref{lindblad_L0}) is to note that $\mathcal{L}_0$ acts on the space of linear operators defined over the joint Hilbert space $\mathcal{H}_{1/2}^{\otimes N}$, where $\mathcal{H}_{1/2}\simeq \mathbb{C}^2$, and as such we can represent an operator as $X=\sum_{\bar n}\mu_{\bar n}\sigma_{\bar n}$, where $\mu_{\bar n} = \frac{1}{2^N}  \text{Tr}[X \sigma_{\bar n}]$ (we use the same notation as in the main text). Then, by linearity,

\begin{equation}
\begin{split}
&\mathcal{L}_0 [X] = \sum_{\bar n} \mu_{\bar n}\, \mathcal{L}_0[ \sigma_{\bar n}] \\
&= \frac{1}{2^N} \sum_{\bar n}  \lambda_{\bar n} \text{Tr}[X \sigma_{\bar n}] \sigma_{\bar n} =: \sum_{\bar n} \lambda_{\bar n} \mathcal{P}_{\bar n}[X]
\end{split}
\end{equation}
where in the second line we have used that $\sigma_{\bar n}$ is an eigenstate of $\mathcal{L}_0$ (with value $\lambda_{\bar n} \in \{0,-2\gamma\}$). In the last step we defined the projector which acts as $\cl{P}_{\bar n}[X] = \frac{1}{2^N}\text{Tr}[X \sigma_{\bar n}] \sigma_{\bar n}$.

We now show $\cl{P}_{\bar n}$ is indeed a genuine projector. We take $X \in L(\mathcal{H}_{1/2}^{\otimes N})$ as above, an arbitrary linear operator over the joint Hilbert space. First,

\begin{equation}
\begin{split}
&\cl{P}_{\bar n}\cl{P}_{\bar m}[X] = \frac{1}{2^N}\sum_{\bar p}\mu_{\bar p} \cl{P}_{\bar n} \text{Tr}[\sigma_{\bar m}\sigma_{\bar p}]\sigma_{\bar m} \\
&=\frac{1}{2^N} \mu_{\bar m}\text{Tr}[\sigma_{\bar m}\sigma_{\bar n}]\sigma_{\bar n} 
 = \delta_{\bar m \bar n}  \mu_{\bar n} \sigma_{\bar n} = \delta_{\bar m \bar n} \cl{P}_{\bar n}[X]
\end{split}
\end{equation}
where we used $\text{Tr}[\sigma_{\bar m}\sigma_{\bar n}] = 2^N \delta_{\bar m \bar n}$. Since $X$ is arbitrary, we have $\cl{P}_{\bar n}\cl{P}_{\bar m} = \delta_{\bar m \bar n}\cl{P}_{\bar n}$.

Second,
\begin{equation}
\begin{split}
&\sum_{\bar m}\cl{P}_{\bar m}[X] = \sum_{\bar m, \bar n} \mu_{\bar n} \cl{P}_{\bar m}[\sigma_{\bar n}] \\
& = \frac{1}{2^N} \sum_{\bar m, \bar n} \mu_{\bar n} \text{Tr}[\sigma_{\bar m}\sigma_{\bar n}]\sigma_{\bar m} = \sum_{\bar n} \mu_{\bar n}\sigma_{\bar n} = X
\end{split}
\end{equation}
so that $\sum_{\bar m}\cl{P}_{\bar m} = \mathbb{I}$.

\subsection{Derivation of $\Lambda(t)$ [Eq.~(\ref{eq-lambda})] \label{sec:deriv}}
We assume $\omega \in \mathbb{R}_{>0}$ defined in the main text is real and non-zero. Apart from steady states, the eigenvalues of $\cl{L}_0$ are $\lambda=-2 \gamma$. Recall we have $\cl{L}_1 \propto \cl{L}_0$ (the same up to a positive constant), and we absorb the (magnitude of the) non-zero eigenvalue of $\cl{L}_1$ in $B$ (so as to avoid introducing a redundant parameter). We compute the $\tilde \Lambda (s)$  (as in Eq.~(\ref{laplace-solution})), for these eigenvalues:
\be
\begin{split}
\tilde \Lambda(s) = \frac{1}{s + 2\gamma + \frac{B^2}{s+\tau_{k}^{-1}}} = \frac{s+\tau_{k}^{-1}}{(s-s_+)(s-s_-)}\\
= \frac{c_+}{s-s_+} + \frac{c_-}{s-s_-}
\end{split}
\ee
where $s_\pm = -\tau^{-1} \pm i\omega$, and $c_\pm = \frac{1}{2}(1\pm i\frac{2\gamma \tau_{k} - 1}{2\omega \tau_{k}})$.

Therefore, $\Lambda(t) = c_+e^{s_+ t}+c_-e^{s_- t}$. We can write $c_\pm = |c|e^{\pm i \phi}$, which gives
\be
\begin{split}
\Lambda(t) = |c|e^{-t/\tau}(e^{i\phi}e^{i\omega t }+e^{-i\phi}e^{-i\omega t })
\\
=2|c|e^{-t/\tau}\cos (\omega t + \phi),
\end{split}
\label{lambda-cosine}
\ee
where $2|c| = \sqrt{1+(2 \gamma \tau_{k} - 1)^2/(2 \omega \tau_{k})^2} = 1/\cos \phi$.

Note, by expanding the cosine function, this can also be written as
\be
\Lambda(t) = e^{-t/\tau}(\cos (\omega t) - \tan \phi \sin (\omega t)),
\label{lambda-full}
\ee
where $\tan \phi = \frac{\gamma - 1/2\tau_{k}}{\omega}$. One can in fact use the form Eq.~(\ref{lambda-full}) to easily derive $\Lambda$ in the limit $\omega \rightarrow 0$, or when $\omega = i|\omega|$.

Note that at times $T=\frac{2}{\omega}\pi n$, and $\frac{2}{\omega}(\pi n - \phi)$ [$n=1,2,\dots$], we have $\Lambda(T) = e^{-T/\tau}$, and therefore the evolution operator is
\be
\Phi_T' = \sum_{\bar n} e^{- \lambda_{\bar n}' T}\cl{P}_{\bar n},
\ee
where $\lambda_{\bar n}' = \frac{1}{2}(\lambda_{\bar n} + 1/\tau_{k})$, where $\lambda_{\bar n} $ is either 0 or $-2\gamma$. We see, that the evolution of this system (for time $T$) is equivalent to evolution under the background channel $\cl{L}_0$ alone, with $\gamma$ replaced by $\frac{1}{2}(\gamma +1/2\tau_{k})$. We demonstrate this in the main text in Fig.~\ref{fidelity-fig}, where we set $\tau_{k} \rightarrow \infty$.

\subsection{Conditions for complete positivity \label{sec:cp}}

For map Eq.~(\ref{newMap}) to be CP, we require $0\le p(t) \le 1$, and therefore $-1 \le \Lambda(t) \le 1, \forall t\ge 0$.  First we consider $\omega \in \mathbb{R}_{>0}$ (see below for the imaginary case) \footnote{Note, if in fact $\omega=0$, one can easily see directly that $|\Lambda(t)| \le 1$.}, and therefore we have $\Lambda(t) = e^{-t/\tau}\cos (\omega t +\phi)/\cos \phi$.

We differentiate this which shows at the turning points, $\hat t$, we have $\cos (\omega \hat t + \phi) = \frac{\pm \tau \omega}{\sqrt{1+(\tau \omega)^2}}$, and therefore
\be
\frac{|\cos (\omega \hat t + \phi)|}{\cos \phi}  = \sqrt{\frac{1+\chi_-^2}{1+\chi_+^2}}<1,
\ee
where $\chi_\pm = \frac{2 \gamma \tau_{k} \pm 1}{2 \omega \tau_{k}}$. Thus, $\Lambda(\hat t) \le 1$. Since also $\Lambda (0)=1,\, \Lambda (\infty) = 0$, it is clear that $|\Lambda(t)| \le 1$ for all parameters, and all $t\ge 0$.

\subsubsection{The case $\omega = i |\omega|$}
\label{sect-app-imaginary}
We define for convenience $\eta = 2\gamma \tau_{k}$.

If $\omega$ is not real, then it is purely imaginary of the form $\omega=i|\omega|$ (this occurs when $|B| < \gamma |1 -1/\eta |$). In this case the analysis is simple since we can see that, from Eq.~(\ref{lambda-full}) above,
\be
\Lambda(t) = e^{-t/\tau}\left[ \cosh |\omega | t + \frac{-\gamma + 1/2\tau_{k}}{|\omega|}\sinh |\omega |t \right],
\ee
and therefore 
\be
|\Lambda (t)| \le \Lambda^*(t) : = e^{-t/\tau}\left[ \cosh |\omega | t + \frac{1}{|\omega| \tau}\sinh |\omega |t \right].
\ee
We see 
\be
\frac{d\Lambda^*}{dt} = e^{-t/\tau}|\omega |\sinh (|\omega |t)\left(1-\frac{1}{|\omega |^2 \tau^2}\right)\le 0,
\ee
where the inequality comes from the observation that $|\omega |^2 = \gamma^2(1 - 1/\eta)^2 -B^2 < \gamma^2( 1+ 1/\eta)^2 = \tau^{-2}$.

Since $\Lambda^*(0)=1$, and is decreasing for all times, it is clear that $\Lambda^*(t) \le 1, \forall t \ge 0$.

\subsection{Modulated decay noise - derivations \label{sec:cosine}}
As described in the main text, we have 
\be
\tilde{k}(s) = B^2 \frac{s + \tau_k^{-1}}{(s + \tau_k^{-1})^2 + \nu^2}. 
\ee
Note, as before, we can absorb any redundant (positive) constants into the definition of $B$.
Therefore, the poles of $\tilde{\Lambda}(s)$ are the roots of
\begin{equation}
\label{poly}
\begin{split}
s^3 + 2s^2(\gamma+\tau_k^{-1}) + s (\nu^2 + B^2 + 4\gamma \tau_k^{-1} + \tau_k^{-2}) \\
 + 2\gamma(\tau_k^{-2}+\nu^2) + B^2 \tau_k^{-1}.
\end{split}
\end{equation}

We take $B^2 = \frac{2}{9}(2\gamma - \tau_k^{-1})^2 + 2 \nu^2$, which means the roots of Eq.~(\ref{poly}) are simply
\begin{equation}
s^* = -\tau^{-1},\, -\tau^{-1} \pm i \omega
\end{equation}
where $\tau^{-1}$ and $\omega$ are given in the main text.

 Therefore, we can write using partial fractions
\be
\tilde{\Lambda}(s) = \frac{c_0}{s + \tau^{-1} } + \frac{c_1}{s + \tau^{-1} - i \omega} + \frac{c_1^*}{s+\tau^{-1} + i\omega},
\ee
for constants
\be
\begin{split}
& c_0 = \frac{1}{9\omega^2}\left((2\gamma - \tau_k^{-1})^2 + 9\nu^2\right), \\
& c_1= \frac{1-c_0}{2\cos \phi}e^{i\phi} \\
& \cos \phi = \frac{1-c_0}{\sqrt{(1-c_0)^2 + \frac{4}{9\omega^2} (2 \gamma - \tau_k^{-1})^2}}.
\end{split}
\ee
Inverting this, one gets
\be
\label{cosineLambda}
\begin{split}
\Lambda(t) = e^{- t / \tau }\left[c_0 + c_1e^{i\omega t} + c_1^* e^{-i \omega t}\right] \\
= e^{- t / \tau }\left[ c_0 + \frac{1-c_0}{\cos \phi}\cos (\omega t + \phi)\right].
\end{split}
\ee

We note that the dynamics generate a genuine quantum map if $|\Lambda(t)| \le 1$, $\forall t \ge 0$. For a given parameter set, one can numerically check this by for example, differentiating Eq.~(\ref{cosineLambda}), to find the first minima and maxima of $\Lambda(t)$ (subsequent minima/maxima will be lower/upper bounded by the first due to the exponential). If at these turning points $|\Lambda(t)| \le 1$, the map is CP for all times.
Note, as per the main text, for a choice of $\nu,\tau_k$, one must also check that setting $\gamma=0$ still generates a CP map. 

Lastly, notice that at times $T = \frac{2}{\omega}\pi n $ , and $\frac{2}{\omega}(\pi n - \phi)$, [$n=1,2,\dots$] then $\Lambda(T) = e^{- T / \tau}$, and the resulting evolution (operator) is equivalent to that under $\cl{L}_0$ alone, with $\gamma$ replaced by $\frac{1}{3}(\gamma + \tau_k^{-1})$. For large $\tau_k$, we can reduce the decay rate by nearly a factor of three (as compared to a factor of two with the purely exponential kernel).

\subsection{Thermal Spectral Theorem \label{app:thermal}}
We provide a reminder of the dynamics for a qubit in a thermal (Markov) environment.

For our dissipative model, let us consider a qubit in a thermal environment, which we describe by the generator $\cl{L}_0 = \cl{L}_- + \cl{L}_+$, where, for $\alpha=\pm$,
\begin{equation}
\label{thermal}
\cl{L}_\alpha [X] = \gamma_\alpha \left( \sigma_\alpha X \sigma_{\alpha}^{\dagger}  -\frac{1}{2}\{\sigma_{\alpha}^{\dagger}\sigma_\alpha,X\}\right),
\end{equation}
with $\sigma_\pm = \frac{1}{2}(\sigma_1 \pm i\sigma_2)$. 


Similar to the Pauli channel, this has no eigen-nilpotents, and can therefore be written $\cl{L}_0  = \sum_{i=0}^3 \lambda_i \cl{P}_i$ (and $\lambda_i \in \mathbb{R}_{\le 0}$).
We can therefore introduce the left $L_i$ and right $R_i$ eigenvectors of $\cl{L}_0$, which form an orthonormal basis (see below).

Defining $\Gamma := \gamma_-+\gamma_+$, $x:=\gamma_+/\gamma_-$,
one can show
\begin{equation}
\label{spectral-thermal}
\cl{P}_i[X] = \text{Tr}[L_i X]R_i
\end{equation}
where
\begin{equation}
\begin{split}
&\{\lambda_i\} = \{0,-\frac{1}{2}\Gamma,-\frac{1}{2}\Gamma,-\Gamma\} \\
& \{R_i\} =\{\sigma_0- \frac{1-x}{1+x}\sigma_3,\sigma_-,\sigma_+,-\frac{1}{2}\sigma_3\}\\
&\{L_i\} = \{\frac{1}{2}\sigma_0,\sigma_+,\sigma_-,\frac{x-1}{x+1}\sigma_0 - \sigma_3)\}.
\end{split}
\end{equation}
One can check orthonormality in the sense $\text{Tr}[L_i R_j] = \delta_{i,j}$.

We use this to study the infinite time dynamics. For any quantum state $\rho$,
we have $\rho_\infty := \Phi_{t \rightarrow \infty} \rho = \cl{P}_0\rho = R_0$.
More concisely,  the unique steady state of the dynamics is
$\rho_\infty = \frac{1}{1+x}(|0\rangle \langle 0| + x|1\rangle \langle 1|)$.

This unique steady state can be written as a thermal (Gibbs) state, where we identify an inverse temperature $\beta$ such that $\rho_\infty = \frac{1}{Z}e^{-\beta H}$, where $H=g \sigma_3$ (and $Z = \text{Tr}[e^{-\beta H}]$), with $x = e^{-\beta g}$.

\end{document}